\RequirePackage{rotating}
\documentclass[fleqn,usenatbib]{mnras}

\usepackage[T1]{fontenc}
\usepackage{ae,aecompl}

\usepackage{graphicx}	
\usepackage{amsmath}	
\usepackage{amssymb}	
\usepackage{pifont}

\newcommand{\glafic}{\texttt{glafic}\ }
\newcommand{\grale}{\texttt{Grale}\ }
\newcommand{\lenstool}{\texttt{Lenstool}\ }
\newcommand{\pixelens}{\texttt{PixeLens}\ }
\newcommand{\wslap}{\texttt{WSLAP+}\ }
\newcommand{\sextractor}{\texttt{SExtractor}\ }
\newcommand{\sep}{\texttt{sep}\ }
\newcommand{\ptmatch}{\texttt{ptmatch}\ }
\newcommand{\eagle}{\textsc{Eagle}\ }

\newcommand{\cmark}{\ding{51}}%
\newcommand{\xmark}{\ding{55}}%
\newcommand{\comm}[1]{{#1}}
\newcommand{\news}[1]{{\textcolor{black}{#1}}}

\title[Resolving the anomalies in Abell 3827]{Much ado about no offset -- Characterising the anomalous multiple-image configuration and the model-driven displacement between light and mass in the multi-plane strong lens Abell 3827}

\author[J.~Lin, J.~Wagner, and R.~E.~Griffiths]{
Joyce Lin$^1$\thanks{E-mail: joycelin2809@gmail.com}, Jenny Wagner$^{2}$\thanks{E-mail: thegravitygrinch@gmail.com}, and Richard E.~Griffiths$^{1,3}$\thanks{E-mail: griff2@hawaii.edu}
\\
$^{1}$Department of Physics, Carnegie Mellon University, 5000 Forbes Avenue, Pittsburgh, PA 15213, USA, \\ 
$^{2}$Bahamas Advanced Study Institute and Conferences, 4A Ocean Heights, Hill View Circle, Stella Maris, Long Island, The Bahamas, \\ 
$^{3}$ Department of Physics \& Astronomy, University of Hawaii at Hilo, 200 W. Kawili St, Hilo, HI 96720, USA
}

\date{Accepted XXX. Received YYY; in original form ZZZ}

\pubyear{2023}

\begin{document}
\label{firstpage}
\pagerange{\pageref{firstpage}--\pageref{lastpage}}
\maketitle

\begin{abstract}
Abell 3827 is a unique galaxy cluster with a dry merger in its core causing a highly-resolved multiple-image configuration of a blue spiral galaxy at $z_\mathrm{s}=1.24$.
The surface brightness profiles of four merging galaxies around $z_\mathrm{d}=0.099$ complicate a clear identification of the number of images and finding corresponding small-scale features across them. 
The entailed controversies about offsets between luminous and dark matter have never been settled and dark-matter characteristics in tension with bounds from complementary probes and simulations seemed necessary to explain this multiple-image configuration.
We resolve these issues with a systematic study of possible feature matchings across all images and their impact on the reconstructed mass density distribution. 
From the local lens properties directly constrained by these feature matchings without imposing any global lens model, we conclude that none of them are consistent with expected local characteristics from standard single-lens-plane lensing, nor can they be motivated by the light distribution in the cluster. 
Inspecting complementary spectroscopic data, we show that all these results originate from an insufficient constraining power of the data and seem to hint at a thick lens and not at exotic forms of dark matter or modified gravity. 
\news{If the thick-lens hypothesis can be corroborated with follow-up multi-plane lens modelling, A3827 suffers from a full three-dimensional degeneracy in the distribution of dark matter because combinations of shearings and scalings in a single lens plane can also be represented by an effective shearing and a rotation caused by multiple lens planes.}
\end{abstract} 



\section{Introduction}
\label{sec:introduction}

\begin{figure*}
\centering
\hspace{2ex} \includegraphics[width=0.41\textwidth]{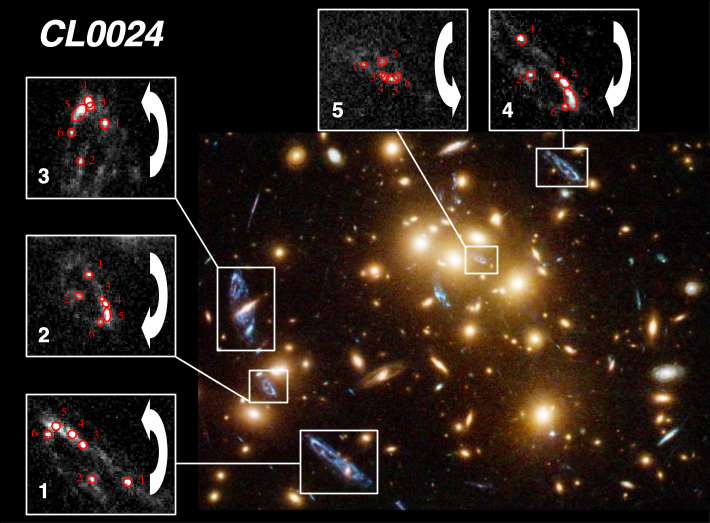}
\hfill
\includegraphics[width=0.41\textwidth]{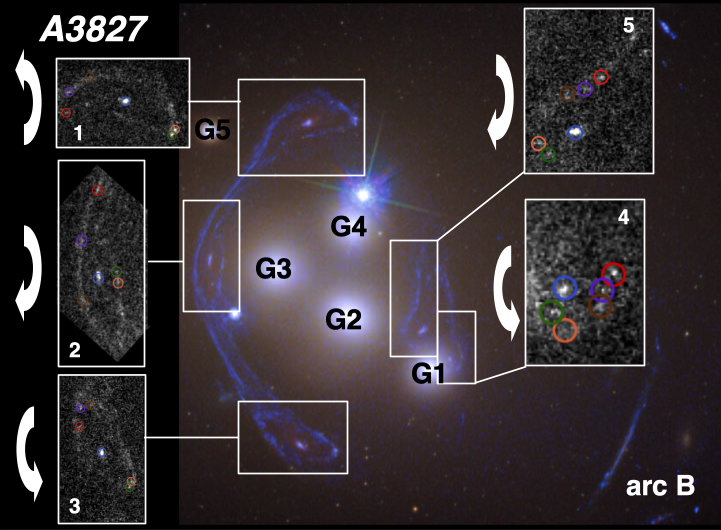} \hspace{2ex}
\caption{Comparison of multiple-image configurations between CL0024 (left) and A3827 (right). 
Relative parities (white arrows) in CL0024 are in agreement with standard cusp and fold configurations in single-plane lensing (see \protect\cite{bib:Wagner_cluster, bib:Lin} for details), while relative orientations in A3827 cannot be brought into agreement with that. 
For instance, the transformation from image~1 to image~2 in A3827 still requires a clockwise rotation of 90 degrees compared to the transformation between image~1 and image~2 in CL0024. 
The central galaxies are labelled G1--4 with G5 being the closest member galaxy outside the multiple-image configuration. 
These galaxies were labelled N1--4 and N6 in \protect\cite{bib:Massey2015}. 
Brightness features in A3827 (coloured circles) are obtained with our persistent-feature extraction pipeline \protect\citep{bib:Lin} as detailed in Section~\ref{sec:brightness_feature}. 
\textit{Image credits: CL0024 adapted from \protect\cite{bib:Wagner_cluster}, A3827 colour image from \protect\cite{bib:Massey2015}, details of multiple images from HST/WFC3 F336W filter band (programme GO-12817).}}
\label{fig:A3827_CL0024}
\end{figure*}

There are only a few galaxy clusters that show clear signs of a merger in their core as convoluted as Abell~3827, A3827 for short. 
Its central part within about 10'' contains a merging asymmetric configuration of four equally bright galaxies with a fifth galaxy close by. 
A blue, rotationally supported background galaxy is multiply imaged into their immediate proximity and the magnification is so strong that all multiple images show a high degree of detailed small-scale features like the spiral arms and bright clumps of potentially star-forming regions apart from the bright central bulge (see Fig.~\ref{fig:A3827_CL0024}, right). 
In addition, there is a small arc, called arc~B and assumed to consist of merged multiple images, at about 20'' south-east of the central galaxies and thus at a larger distance to the cluster centre than the highly detailed multiple images. 

Since the discovery of the strong lensing effect in A3827 by \cite{bib:Carrasco}, the cluster has been investigated many times due to the multiple-image configuration being at odds with those theoretically expected in relaxed, stable strong lensing configurations, as detailed in \cite{bib:Petters}, and actually observed cases like the five-image configuration in CL0024 (see \cite{bib:Wagner_cluster} for a detailed discussion) or the triple-image configurations in SDSS J223010.47-081017.8, also called Hamilton's Object, (see \cite{bib:Griffiths} and \cite{bib:Lin} for details) and in MACS J1149+2223\footnote{part of the Hubble Space Telescope (HST) Frontier Fields \url{https://archive.stsci.edu/prepds/frontier/}}, which used to be the most convoluted cluster until 2009, as noted by \cite{bib:Smith3} (see, among many others, \cite{bib:Kelly3} for a recent analysis of this cluster). 
Unlike statistics based on $N$-body simulated clusters and the vast majority of the observed ones suggest \citep{bib:Wagner7}, A3827 contains one of the few multiple-image configurations on galaxy-cluster scale for which the local lens properties vary over the area covered by the multiple images.

\begin{table*}
\caption{Synopsis of previous reconstructions of the light-deflecting mass density in A3827 separated by the identification of brightness features in the multiple images (horizontal lines). First column: reference; second column: assumed redshift for source galaxy; third column: number of multiple images used as constraints with central images noted separately; fourth column: total number of all brightness features with differing ones mentioned separately; fifth column: \cmark \ if arc~B was used as additional constraint, \xmark \ if not; sixth column: lens model LM used with abbreviations GL=\glafic \citep{bib:Oguri2}, GR=\grale \citep{bib:Liesenborgs_grale, bib:Liesenborgs_grale2}, LT= \lenstool \citep{bib:Jullo_lenstool}, LS=\lenstool with skewed potential \citep{bib:Taylor}, PL=\pixelens \citep{bib:Saha, bib:Coles}, \wslap=WS \citep{bib:Diego, bib:Diego2}; seventh column: offset $\Delta_\mathrm{LTM}$ between centre of light and centre of mass of galaxy~G1.}
\label{tab:related_work}		
\begin{center}
\vspace{-1.5ex}
\begin{tabular}{r|c|l|l|c|c|c|l}
\hline
\noalign{\smallskip}
Reference & $z_\mathrm{s}$ & $n_\mathrm{MI}$ &  $n_\mathrm{f}$ & B? & LM & $\Delta_\mathrm{LTM} \, \left[ \mbox{kpc} \right]$ & Comments \\
\noalign{\smallskip}
\hline
\noalign{\smallskip}
\cite{bib:Carrasco}       & 0.20 & 4 & 9 & \cmark & LT & -- &  Lens model too simple to constrain any offset \\ 
\cite{bib:Williams}        & 0.20 & 4 & 9+1 & \xmark & PL & 6 & Added a visually matching feature in image~2 \\
\cite{bib:Mohammed}  & 0.20  & 4 & 9 & \cmark & GR & 2.1 & Compared offsets in A3827, A2218, A1689 \\
\hline
\noalign{\smallskip}
\cite{bib:Massey2015} & 1.24 & 6 & 30 &\xmark & LT \& GR & 1.62 & Tested alternative image matching \\
\cite{bib:Taylor}            & 1.24 & 6+2 & 30+2 & \xmark & LS & 1.40 & $\Delta_\mathrm{LTM}=1.53~\mbox{kpc}$ with unskewed potential\\ 
\hline
\noalign{\smallskip}
\cite{bib:Massey2018} & 1.24 & 5+2 & 40 & \xmark & LS \& GR & 0.54 & No signif.~offset for any of the four galaxies\\
\hline
\noalign{\smallskip}
\cite{bib:Chen}             & 1.24 & 5+2 & 39 & \xmark & WS \& GL & -- & Indirectly tested alternative gravity (MOND)\\
\hline                              
\end{tabular}
\end{center}
\end{table*}

All related work analysing the multiple-image configuration in A3827 employed single-lens-plane lens modelling reconstructions of the light-deflecting mass density around the central 10'' including the four central galaxies G1--4 and G5 as soon as the latter was recognised as a galaxy \citep{bib:Massey2015}.
Some analyses also included an extended catalogue of distant member galaxies but without significant changes in the resulting mass density reconstructions \citep{bib:Taylor}.  
As constraints, four multiple images of the background galaxy were used first \citep{bib:Carrasco, bib:Williams, bib:Mohammed}, until it became clear that the fourth image is a combination of at least two multiple images and two additional, unresolved central images were spectroscopically confirmed \citep{bib:Massey2015, bib:Massey2018, bib:Chen}.
\cite{bib:Massey2015} and \cite{bib:Taylor} split the fourth multiple image into three separate ones, such that \cite{bib:Massey2015} uses a total of six multiple images and \cite{bib:Taylor} assume a total of eight multiple images, after the discovery of the two central images. 
\cite{bib:Massey2018} and \cite{bib:Chen} split the fourth image only into two and both use the two central images, such that their total amount of multiple images is seven. 
Furthermore, \cite{bib:Carrasco, bib:Williams}, and \cite{bib:Mohammed} placed the background source at redshift $z_\mathrm{s}=0.20$ as determined by \cite{bib:Carrasco}, while follow-up observations detailed in \cite{bib:Massey2015} revealed that the source actually lies at $z_\mathrm{s}=1.24$ along the line of sight. 

The brightness features within the multiple images serving as constraints for the lens models were extracted by \sextractor \citep{bib:Bertin} or by visual inspection in all of the works only after preprocessing to subtract the surface brightness profiles of the bright central galaxies and the spikes of two nearby stars.  
Subsequently, the matching of corresponding features across all images was performed according to visual similarity and only a handful of variations were tested when comparing the matchings of \cite{bib:Massey2015, bib:Taylor, bib:Massey2018}, and \cite{bib:Chen}. 

As lens models, both parametric and free-form models were employed. 
Irrespective of the identified brightness features, their matching, including arc~B as a constraint from source at  different redshift ($z_\mathrm{B}=0.41$) or the lens model used, the works of \cite{bib:Williams, bib:Mohammed, bib:Massey2015, bib:Taylor}, and \cite{bib:Chen} concluded that light does not seem to trace mass in this cluster.
The centre of light of at least one of the bright galaxies, G1, did not coincide with the centre of mass for this galaxy as inferred from their lens-model reconstructions at about 3-$\sigma$ significance. 
Only \cite{bib:Massey2018} arrived at the contrary conclusion after refining the positions of the small-scale features within the multiple images which were used for the modelling of the light-deflecting mass density. 
Yet, two years later, \cite{bib:Chen} challenged this conclusion again on cluster scale with their mass density reconstruction because it required dark matter not to trace the luminous matter.
The latter was modelled as the member galaxies, an intra-cluster stellar part, and a gas component. 

To summarise the status-quo, Table~\ref{tab:related_work} shows the main characteristics and results of all previous investigations. 
In addition, Fig.~\ref{fig:A3827_CL0024} compares the multiple-image configurations of A3827 and CL0024 to highlight the differences between standard lens mappings, represented by CL0024 (left), and the anomalous one observed in A3827 (right). 
A nomenclature for this paper is also introduced and brought into accordance with previous works. 

In this paper, we aim to tackle several still unanswered questions for this cluster and organise the paper accordingly as follows: 
Given the low signal-to-noise environment of the bright cluster galaxies and the foreground stellar contamination, we investigate where we can find robustly identifiable features with a signal strength of at least 3-$\sigma$ above the local noise level.
To answer this question, we apply our recently developed image-processing pipeline to robustly extract persistent brightness features from the multiple images \citep{bib:Lin} to the F336W filter band observation from \cite{bib:Massey2015} without any preprocessing as performed in previous works.
It is possible to forgo the latter because all multiple images are clearly visible in this filter band, while the remainder of the available HST observations in F160W, F814W, and F606W only show very small parts of the outer images, devouring the three central ones completely. 
Hence, we cannot test the assumption that lensing is wavelength independent, as supported by our multi-wavelength analyses of CL0024 and Hamilton's Object \citep{bib:Lin} for A3827 but have to rely on its validity for this case. 
Instead, we can test the robustness of feature extraction without and after a preprocessing that introduces model assumptions about the brightness profiles of the lensing galaxies.
Details about the pipeline and the resulting features extractable out of the five multiple images in A3827 can be found in Section~\ref{sec:brightness_feature}. 

The second question to be tackled is the one about matching brightness features across multiple images. 
Due to partial occlusion, the low signal-to-noise-level, and the large magnification of the multiple images there may not be a unique way of matching the extracted features across all images. 
Features can be missing for small images or images in a bright environment (like image~4) or additional features can be visible in large images that are less affected by stray light (like image~1), both just based on the varying magnification for the images. 
Besides, misidentifications due to features highlighted by microlensing or features attenuated by dust can occur. 
The ambiguity occurring in the feature matching for Hamilton's Object was easy to identify and its direct impact on the reduced shear components of this multiple image could be immediately tracked in the determination of the local lens properties, as detailed in \cite{bib:Lin}. 
Contrary to this clear case, the feature matching in A3827 is more intricate and we systematically investigate the possibilities and develop an algorithm to find the optimum matching for our lens-model-independent reconstruction of local lens properties, as summarised in \cite{bib:Wagner_summary}, in Section~\ref{sec:optimum_feature}. 
Based on the insights gained from this analysis, we will evaluate the quality of the features found in \cite{bib:Massey2018} and \cite{bib:Chen} as well. 
In principle, the amount of features that can be matched across all five multiple images is large enough to even investigate the variation of local lens properties across the area covered by the multiple images as an interesting side product. 
A3827 contains very extended multiple images compared to the approximate radius of their Einstein ring, such that it is a rare example for which such an analysis is reasonable.

The third question we will tackle concerns the constraining power of the observable brightness features to significantly claim the existence of an offset between light and mass in Section~\ref{sec:comparative_discussion}. 
As derived in \cite{bib:Tessore, bib:Wagner_summary}, and \cite{bib:Wagner7}, we can only constrain the differential, locally distorting properties of the light-deflecting mass density within the area covered by the multiple images from the observables.
Any properties of the lens at other locations are dependent on a lens model, which acts as an extrapolation scheme from the data-inferred properties into regions devoid of data. 
Yet, since galaxy G1 is so close to images~4 and 5, it is possible that the extrapolation via a lens model is small enough to put constraints on a potential offset.
\news{We will briefly comment on that in Section~\ref{sec:comparative_discussion} based on our lens-model-independent results and the free-form lens reconstructions obtained by \grale \citep{bib:Liesenborgs_grale, bib:Liesenborgs_grale2} in previous works. 
As detailed in \cite{bib:Wagner_cluster}, the confidence bounds obtained from free-form lens reconstructions represent the remaining freedom of the lens model not constrained by the observables and therefore, free-form models are ideal to determine whether the observed data has sufficient constraining power to detect any offset. The confidence bounds of parametric lens models, on the other hand, show how well a certain assumed mass density distribution fits to the observed data. They thus yield a complementary information with respect to the free-form approaches.} 
Using only the multiple images from a single background source in most previous work, we will also investigate the influence of the still unbroken degeneracies like the global and local versions of the mass-sheet degeneracies (see \cite{bib:Falco, bib:Gorenstein, bib:Wagner4, bib:Wagner6} for further details on the general matter of degeneracies and \cite{bib:Liesenborgs1, bib:Wagner_frb, bib:Liesenborgs_grale2} for an encompassing characterisation of the degeneracies occurring in \grale as a particular lens modelling approach).

Having answered these  questions, we summarise all our findings in a consistent picture in Section~\ref{sec:comparative_discussion}. 
In this context, we will also investigate why the multiple-image configuration in A3827 deviates from the standard ones, like the five images in CL0024 (Fig.~\ref{fig:A3827_CL0024}, left), in Section~\ref{sec:conclusion}. 
We will show that the single-lens-plane approach may need an extension along the line of sight due to the immediate proximity of the lensing cluster with redshift $z_\mathrm{d}=0.099$ away from us.
The spectroscopic analysis of the cluster member galaxies of \cite{bib:Carrasco} arrived at a bi-modal distribution of galaxy velocities, thus supporting at least two different lens planes for A3827.
As a consequence, casting the thick lens as a thin one projecting the mass density into a single lens plane, the necessary image rotations to capture the multi-plane nature of this image configuration are reconstructed as a sequence of shears and scalings in lens-model-based single-lens-plane approaches.
This entails ``phantom mass agglomerations'' in the lens reconstruction which are located in regions where no luminous counterpart can be found and which are hardly constrained by observables, as could be the case for A3827.
We also comment on the constraints of deviations from the cold-dark-matter paradigm \citep{bib:Bertone}, which could have been implied by an offset between the centre of mass and the centre of light. 
As lens-model-based reconstructions of the light-deflecting mass density can create phantom dark-matter clumps which generate the shear components necessary to mimic a rotation, it is questionable to infer characteristics about the nature of dark matter from such lens-model-dependent reconstructions without being aware of their limits and behaviour when standard approximations, like dynamical equilibrium, fail to hold. 
\news{Section~\ref{sec:outlook} gives suggestions for complementary data that can be added to the strong-lensing observables to alleviate this degeneracy between single- and multi-plane lensing configurations.}

\vspace{-2ex}
\section{Brightness feature extraction}
\label{sec:brightness_feature}

As stated in Section~\ref{sec:introduction}, we use the single 5871s exposure in the F336W UV filter band observation taken with HST/WFC3 in the programme GO-12817 during August 2013, which is publicly available in the Hubble Legacy Archive (HLA) in order to extract the brightness features required for our lens-model-independent inference of local lens properties. 
Given that the galaxies G1--4 outshine parts of the cluster centre in other filter bands and their resolution is insufficient to identify single brightness features (see Table~\ref{tab:filter_bands}), the observable features visible in F336W drive the lens reconstruction. 

All details about our feature-extraction method can be found in \cite{bib:Lin}. 
In Section~\ref{sec:feature_extraction}, we present the results of the pipeline for A3827.
Section~\ref{sec:microlensing_attenuation} subsequently evaluates whether the features found by our method are microlensed or attenuated, as can be inferred from a comparison of observables in the F336W filter band with other filter bands available and over observation epochs (as far as this is possible). 
We also investigate whether some of them could be source-intrinsic transients which are subject to time-delay differences between their occurrence in the multiple images. 
At the end, we summarise the persistent features as retrieved from our pipeline in Section~\ref{sec:summary_persistent}. 

\begin{figure*}
\centering
\includegraphics[width=0.89\textwidth]{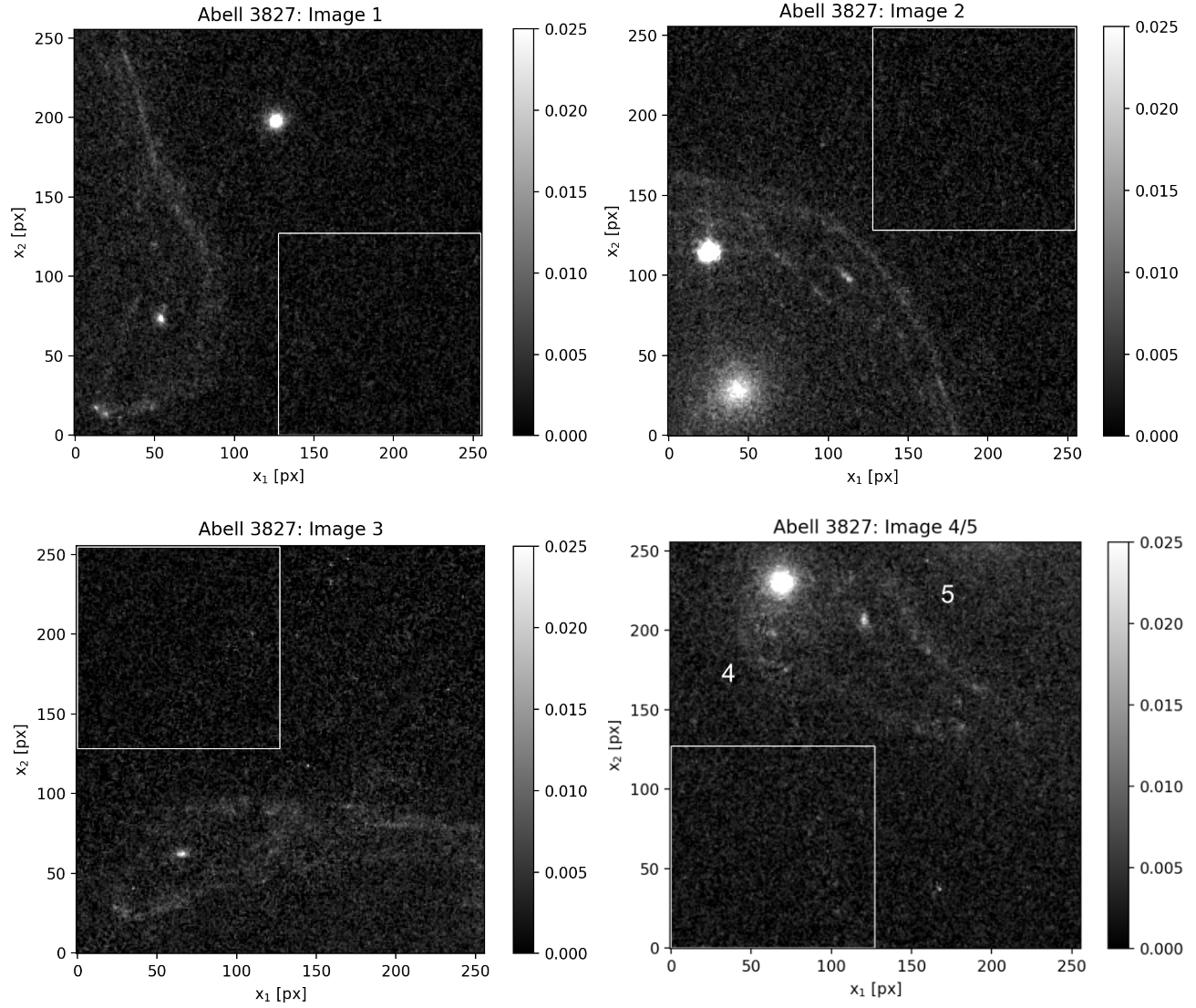}
\caption{Original data used in Section~\ref{sec:brightness_feature} to perform the robust feature extraction: Images~1–5 in the cluster-scale lens Abell 3827 in the HST/WFC3 F336W filter band in $256\times256$ boxes including $128\times128$ patches (white squares) for background estimation. Images~4  (left of G1) and 5 (right of G1) are close together and smaller and therefore in a common box. }
\label{fig:A3827_sep}
\end{figure*}

\vspace{-1.5ex}
\subsection{Feature extraction with our persistence pipeline}
\label{sec:feature_extraction}

In order not to introduce any model-dependent biases like making prior assumptions about the surface brightness profiles of the five central galaxies, the observation in the F336W filter band was not altered or processed in any way before our analysis. 
The first step in our image-processing pipeline is the calculation of the local background intensity, $\mu_\mathrm{bg}$, and its root-mean-square variation, $\sigma_\mathrm{I}$, in the close vicinity of each multiple image to determine the noise level in the observation. 
We use the Python library for Source Extraction and Photometry, \sep\!, \citep{bib:Barbary} to do so. 

Images~1, 2, and 3 each cover an area on the sky that is approximately four times larger than their counterparts in CL0024 and Hamilton's Object, which were used to test the image-processing pipeline detailed in \cite{bib:Lin}. 
Due to the increased image size, the box size that contains the multiple images and a representative sample to characterise the local background is four times larger, $256\times256$ pixels. 
Accordingly, the size of the box containing the sample of the local background is enlarged four times as well to $128\times128$ pixels. 
Compared to the previously analysed clusters, intensity variations are much larger due to the very bright galaxies G1--4. 
Therefore, determining a precise local estimate of the background intensity value and the noise level is more important than it used to be in CL0024 or for Hamilton's Object. 
As can be seen in Fig.~\ref{fig:A3827_sep}, the location of the boxes to constrain $(\mu_\mathrm{bg}, \sigma_\mathrm{I})$ (white squares) do not contain any bright foreground objects. 
Their locations are carefully chosen not to be close to the bright central galaxies, either. 

While this approach not subtracting the brightness profiles of G1--5 may lead to some features being artificially boosted above the detection threshold by the impact of G1--5, Fig.~\ref{fig:A3827_sep} shows that this impact is at least as small as the impact of a brightness profile with a narrow width. 
Our approach refrains from extracting features out of the areas occluded by the bright galaxies and it is less biasing than the approach pursued in \cite{bib:Chen} because they \emph{heuristically} adjust the brightness profile map of the galaxies from a different waveband to be subtracted from F336W. 

Next, we create persistence diagrams for candidate brightness features, called objects, by thresholding the background-subtracted intensity map of F336W at $\mu=\mu_\mathrm{bg}+n_\mathrm{I} \sigma_\mathrm{I}$ for a range of different $n_\mathrm{I} \ge 3$. 
Thus, the minimum intensity defining an object is 3-$\sigma$ above noise level. 

\begin{figure*}
\centering
\hspace{2ex}\includegraphics[width=0.4\textwidth]{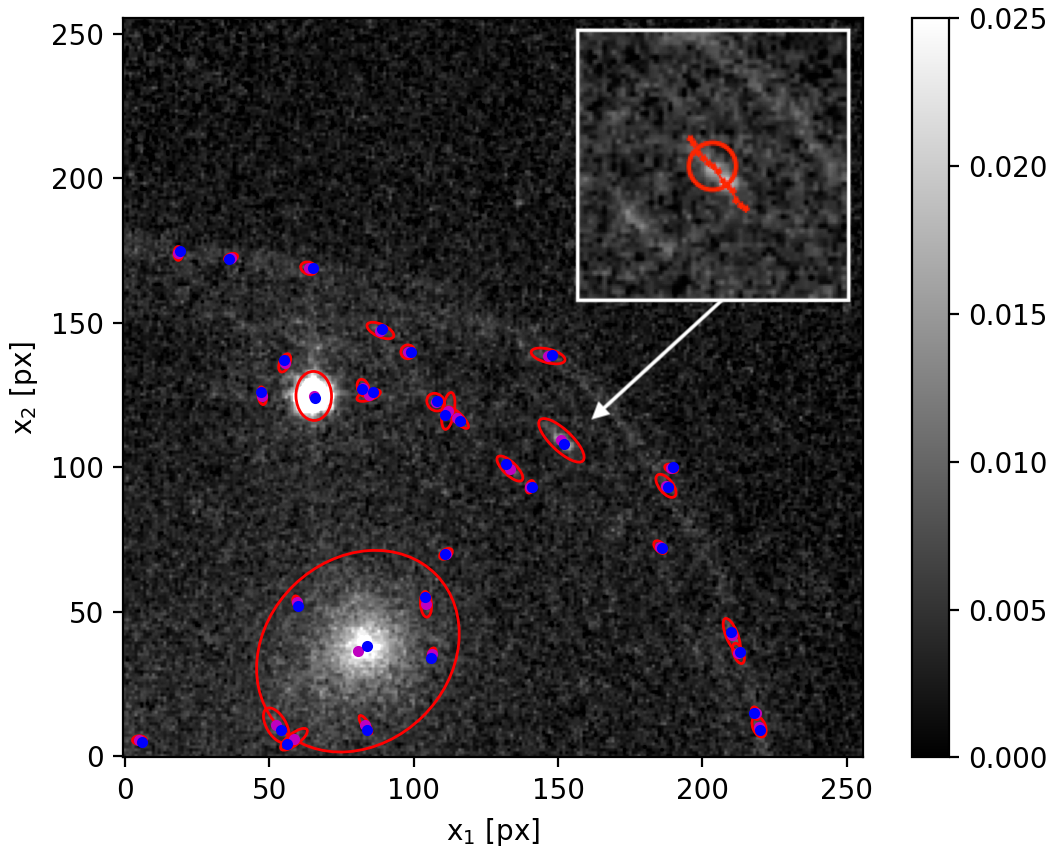} \hfill
\includegraphics[width=0.4\textwidth]{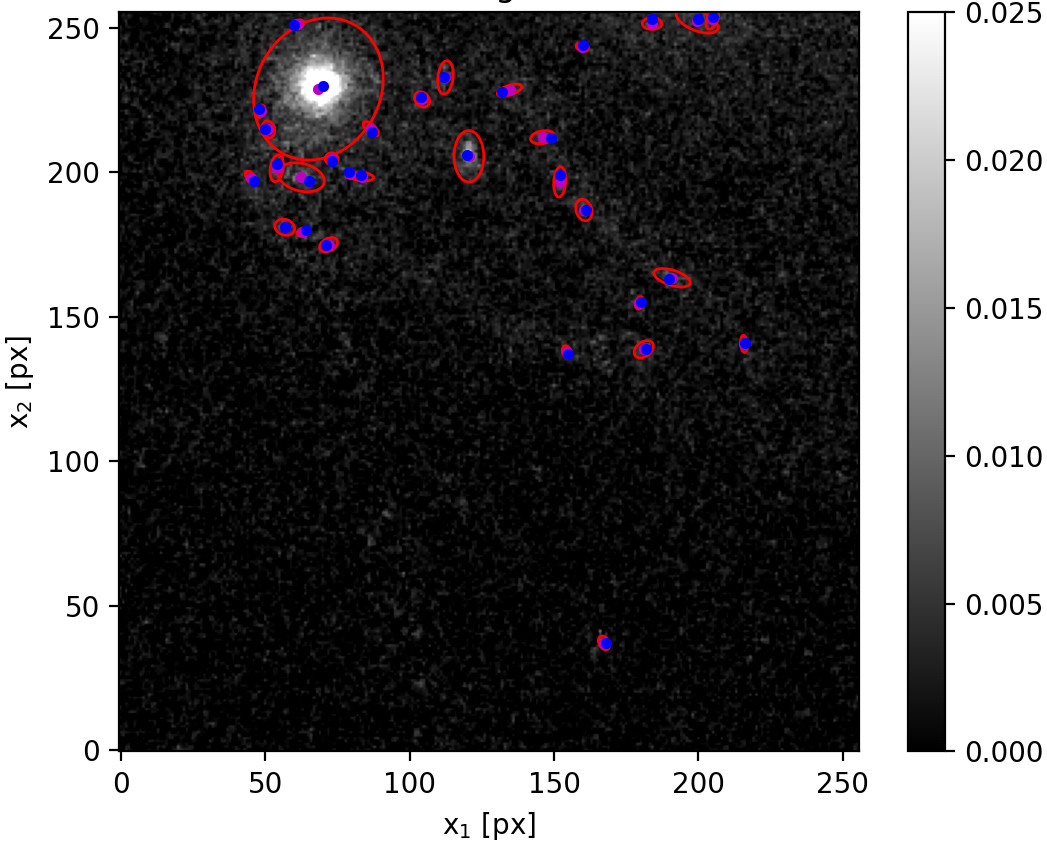} \hspace{2ex}
\caption{Objects detected above $n_\mathrm{I}=3$ in image~2 (left) and images~4 and 5 (right), characterised by the locations of the peak intensity (blue dots), their centre of light (magenta dots), and their extension in terms of the intensity quadrupole (red ellipses). The central bulge object in image~2 is not unimodal and possible feature locations can be found along an entire line (red dots) shown in the inlay (white square in the left plot). The final position of the feature is marked by a red circle.}
\label{fig:image245_objects}
\end{figure*}

Fig.~\ref{fig:image245_objects} (left) shows the locations of the peak intensity (blue dot), the centre of light (magenta dot) and the extension of each object via its quadrupole (red ellipse) for image~2 for $n_\mathrm{I}=3$. 
For comparison, we also show the same plot for images~4 and 5 in Fig.~\ref{fig:image245_objects} (right) to demonstrate the impact of G1 on the feature extraction. 
The plots for the other images are in Appendix~\ref{app:brightness_feature}.
As expected, the density of objects increases in the vicinity of the bright foreground galaxies.
Nevertheless, there are also objects above the minimum threshold on the darker sides of the multiple images.  
The low signal-to-noise environment compared to the previously analysed galaxy clusters can be inferred from the offsets between the centres of light and the positions of the peak intensity, as detailed theoretically in \cite{bib:Lin}. 

Due to the large amount of objects at low thresholds, the persistence graphs list a lot of entries and are therefore moved to Appendix~\ref{app:brightness_feature}. 
As motivated in \cite{bib:Lin}, the longer an object persists over increasing $n_\mathrm{I}$, the more likely that it is a reliable signal and not an artefact of noise. 
The strong decrease of persisting features for thresholds $n_\mathrm{I}>3$ underlines the low signal-to-noise of the observation.

While, in general, the bright central bulges are more extended than all other brightness features in all multiple images, the central bulge in image~2 is extremely extended into a bright line, even splitting into a bi-modal intensity distribution at higher $n_\mathrm{I}$, see Fig.~\ref{fig:A3827_sep}. 
It is the largest and the most persistent object (as a whole) in the entire analysis. 
Consequently, the uncertainty in the position of the central bulges is larger than for the other features and, as we will further detail in Section~\ref{sec:optimum_feature} using the central bulge as a feature in the matching process deteriorates the quality of the inferred local lens properties. 
To analyse the impact of the most uncertain position of the central bulge in image~2, we allow its location to be anywhere on the red line shown in the inlay (white square) in Fig.~\ref{fig:image245_objects} (left) to find its location corresponding to those in the other multiple images. 
 
Besides this, not every object persistent over many thresholds in these diagrams is a brightness feature within a multiple image because some of these objects also belong to the bright foreground stars or galaxies located in the same region on the sky. 
\cite{bib:Massey2018} also note that the feature extraction is additionally contaminated by globular clusters in A3827 having similar characteristics as the brightness features within the multiple images. 
These foreground entities have to be ignored when identifying the brightness features to be matched across all images. 
Support for an object to be a genuine feature instead of a foreground artefact can be found in the occurrence of similar features in the corresponding parts of other multiple images. 
For the globular clusters between images~4 and 5 referred to in \cite{bib:Massey2018}, this issue is resolved in an even clearer way because these clusters are below the 3-$\sigma$ threshold and not recognised as objects in our image-processing pipeline. 

To select the objects that are corresponding brightness features across all multiple images in A3827, we develop a new systematic testing routine, outlined in Section~\ref{sec:optimum_feature}. 
As a result, we obtain a self-consistent set of features and the most likely local lens properties constrained by these features as summarised in Section~\ref{sec:summary_persistent}. 

\vspace{-2ex}
\subsection{Microlensing, attenuation, and transients in the source}
\label{sec:microlensing_attenuation}

Visually matching objects with similar surface brightness profiles as corresponding features across multiple images, as has been done with the two right-most features in image~1 and the features at the bottom of image~3 (orange and green circles in Fig.~\ref{fig:A3827_CL0024}), may misidentify correspondences due to serendipitous microlensing. 
In addition, attenuation of features due to dust absorption or due to partial occlusion by bright foreground objects can also introduce confusions in the feature matchings.
A third issue leading to misidentifications are source-intrinsic transients which suffer from time-delay differences in their appearance at the multiple-image locations.
So far, none of these effects and their entailed impact on the mass-density reconstruction have been discussed for the multiple images in A3827.
 
As already detailed in \cite{bib:Lin}, we identify microlensing events and source-intrinsic transients by tracking the differences in the surface brightness profiles of the objects across observation epochs. 
Hints for attenuation due to dust can be found when tracking differences of the surface brightness profiles across multiple filter bands. 

While these investigations can work well for the image configuration in CL0024 and Hamilton's Object, they are only of limited use in A3827, as we will show. 
An encompassing observation using integrated field spectroscopy, as performed by \cite{bib:Massey2018}, proved to be of great help in the identification of the central bulges in images~4 and 5 and to discover the two featureless, central images~6 and 7 not shown in Fig.~\ref{fig:A3827_CL0024}. 
However, an increase in the resolution of the results obtained is still necessary to investigate the configuration at the level of individual features.

To distinguish source-intrinsic transients from microlensing, a lens-model-based mass density reconstruction needs to set up estimates for the time-delay differences expected in A3827. 
However, the constraining power of multiple images from a single source at one redshift is not very high due to the formalism-intrinsic and model-based degeneracies \citep{bib:Wagner_frb}.
Fortunately, as we detail below, there is no need to perform such an analysis for A3827 as the available data can rule out the hypothesis for the two features of interest in the orange and green circles in Fig.~\ref{fig:A3827_CL0024}. 

\begin{table*}
\caption{Observations available for A3827. First column: telescope and filter band of the observation; second column: reference; third column: observation date by month and year; fourth column: \cmark\, if the raw data (R) is available for our analysis, \xmark\, if not; fifth column: source used, figure from reference or data base where raw data is stored; sixth column: details of observable features as marked in Fig.~\ref{fig:all_features} (left), B~=~bulges, FG~=~feature groups.}
\label{tab:filter_bands}		
\begin{center}
\vspace{-1.5ex}
\begin{tabular}{l|c|c|c|c|l}
\hline
\noalign{\smallskip}
Data & Reference & Obs.~date & R? & Source & Observables \\
\noalign{\smallskip}
\hline
\noalign{\smallskip}
GMOS r'-, g'-, i'-bands & \cite{bib:Carrasco} & Nov.~2007 & \xmark & Fig.~1 & Images~1--4 (B, FG)  \\
\hline
HST F336W filter band & \cite{bib:Massey2015} & Aug.~2013 & \cmark & HLA & Images~1--5 (all features) \\
HST F160W filter band & \cite{bib:Massey2015} & Aug.~2013 & \cmark & HLA & Images~1--4 (B), images~1--5 (FG) \\
\hline
HST F606W filter band & \cite{bib:Massey2015} & Oct.~2013 & \cmark & HLA & Images~1, 3, image~2 (red, purple, brown F) \\
HST F814W filter band & \cite{bib:Massey2015} & Oct.~2013 & \cmark & HLA & Images~1, 3, image~2 (red, purple, brown F)\\
\hline
MUSE $\left[\mbox{OII}\right]$ narrow-band & \cite{bib:Massey2015} & Dec.~2014 & \xmark & Fig.~3 & Images~1--5 (B, FG) \\
MUSE $\left[\mbox{OII}\right]$ narrow-band & \cite{bib:Massey2018} & Jun.~2016 & \xmark & Fig.~3 & Images~1--5  (B, FG) \\
\hline
ALMA CO(2-1)-transition & \cite{bib:Massey2018} & Oct.~2016 & \xmark & Fig.~2 & Images~1--5  (B) \\    
\hline                           
\end{tabular}
\end{center}
\end{table*}

Table~\ref{tab:filter_bands} gives a summary of available data.  
To investigate whether the features in the green and orange circles in images~1 and 3, as shown in Fig.~\ref{fig:A3827_CL0024}, are serendipitous events of microlensing, we can thus compare the observations of the Gemini Multi-Object Spectrograph (GMOS) from 2007 with the ones taken by HST in 2013 and subsequently by the Multi-Unit Spectroscopic Explorer (MUSE). 
Since these features, individually or at least as a group in the lower resolution data, persist across all observations over a time span of nine years, serendipitous microlensing or a source-intrinsic transient event seem highly unlikely to explain the similarity in surface brightness and relative positions in these two images.
Consequently, matching these features as done in all previous works seems reasonable, even though this matching clearly breaks the standard relative orientation between these two images in a cusp configuration (like in CL0024, see Fig.~\ref{fig:A3827_CL0024}). 
A confusion with foreground objects like bright stars in intra-cluster gas clouds is equally unlikely, as the two feature groups also appear in the MUSE $\left[\mbox{OII}\right]$ narrow-band observations, which are transformed to the rest-frame of the source.
 
\begin{figure*}
	\centering
	\vspace{-2ex}
	\includegraphics[width=0.93\textwidth]{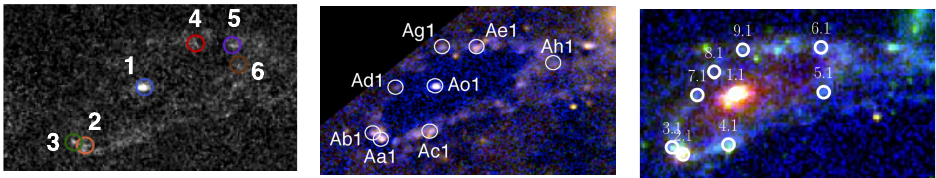}
	\caption{\news{Comparison of features between our best-fit features of Table~\ref{tab:all_features} (left), the features identified by \protect\cite{bib:Massey2018} (centre), and the ones by \protect\cite{bib:Chen} (right) for image~3. Features ``o,a,c'' of \protect\cite{bib:Massey2018} correspond to features~1, 2, 4 of \protect\cite{bib:Chen} and are used to obtain local lens properties from \ptmatch in Table~\ref{tab:comparison_reconstructions}. 
\textit{Image credits: Detail of image~3 from \protect\cite{bib:Massey2018} Fig.~1 (centre) and from \protect\cite{bib:Chen} Fig.~A1 (right).}}}
	\label{fig:feature_comp}
\end{figure*}

\subsection{Summary of persistent features}
\label{sec:summary_persistent}

\begin{table*}
\caption{$(\mbox{RA}, \mbox{Dec})$-coordinates of optimum features for all images~1--5 marked in Fig.~\ref{fig:all_features} (left). All distances in arcseconds relative to the centre of light of G1 at $(330.47518~\mbox{deg},-59.945985~\mbox{deg})$, uncertainties in the positions are $\sigma_x = 0.2"$ for feature~1 (bulge, blue circle in Fig.~\ref{fig:all_features}) and $\sigma_x = 0.1"$ for all other features.}
\label{tab:all_features}		
\begin{center}
\begin{tabular}{l|l|l|l|l|l}
\hline
\noalign{\smallskip}
Feature & Image~1 & Image~2 & Image~3 & Image~4 & Image~5 \\
\noalign{\smallskip}
\hline
\noalign{\smallskip}
1 (blue, bulge) & $(-15.4'',-2.4'')$ & $(-12.1'',6.8'')$ & $(-0.5'',8.6'')$ & $(0.4'',-1.3'')$ & $(-1.8'', -1.0'')$ \\
2 (orange)       & $(-14.0'',-4.8'')$ & $(-11.4'',6.4'')$ & $(\phantom{-}1.0'',7.2'')$ & $(1.2'',-1.3'')$ & $(-1.4'',\phantom{-}0.1'')$ \\
3 (green)         & $(-13.8'',-4.6'')$ & $(-11.8'',6.1'')$ & $(\phantom{-}1.3'',7.3'')$ & $(0.9'',-1.1'')$ & $(-1.1'',-0.2'')$ \\
4 (red)             & $(-14.8'',\phantom{-}0.4'')$ & $(-14.5'',4.1'')$ & $(-1.7'',9.7'')$ & $(0.2'',-2.2'')$ & $(-3.4'',-1.7'')$ \\
5 (purple)        & $(-15.8'',\phantom{-}0.2'')$ & $(-13.6'',6.1'')$ & $(-2.6'',9.7'')$ & $(0.5'',-2.0'')$ & $(-3.0'',-1.2'')$ \\
6 (brown)        & $(-16.5'',-0.7'')$ & $(-12.1'',7.9'')$ & $(-2.7'',9.2'')$ & $(0.7'',-1.9'')$ & $(-2.9'',-0.7'')$ \\          
\hline              
\end{tabular}
\end{center}
\end{table*}

In this section, we applied our image-processing pipeline as developed in \cite{bib:Lin} to the HST F336W filter band of \cite{bib:Massey2015} without any preprocessing to subtract the surface brightness profiles of the central galaxies or the foreground stars.
It discards objects with intensities less than 3-$\sigma$ above local noise level and thereby resolves possible confusions between genuine brightness features in the multiple images and objects that may be artefacts of the model-based preprocessing or that could be actual foreground objects with low significance like the line of globular clusters in front of images~4 and 5 \citep{bib:Massey2018}. 
\news{This is of great importance because previous works based on radio-band data \citep{bib:Wagner_quasar, bib:Ivison} already showed that artefacts can lead to inconsistencies or even strongly biased interpretations of the observables. 
The same argumentation applies to data with low signal-to-noise levels in other filter bands, which is why it can also affect the mass-density reconstruction of A3827.}

Low signal-to-noise ratios are clearly visible due to the differences observed in peak versus centre-of-light locations for the objects and the steep decrease in their amount when increasing the intensity threshold $n_\mathrm{I}$ in the persistence diagrams (see Fig.~\ref{fig:image245_objects} and Appendix~\ref{app:brightness_feature}). 
Object detections in the vicinity of bright foreground objects may be biased towards higher intensities and larger numbers, but are not dependent on any prior assumption about the distribution of the foreground light as all previous works have assumed. 

The hypothesis that the identifications of the features in the green and orange circles in images~1 and 3 (see Fig.~\ref{fig:A3827_CL0024}) could be misidentifications caused by microlensing or a serendipitous alignment of bright foreground objects were rejected. 
Consequently, the anomalous configuration of the multiple images compared to standard ones is corroborated. 

Due to high complexity of the multiple-image configuration in A3827, the set of all objects for each multiple image cannot be clearly mapped to the ones of the other multiple images in contrast to the previous configurations analysed in CL0024 and Hamilton's Object. 
As detailed in Section~\ref{sec:optimum_feature}, we set up a new systematic testing routine to select a set of brightness features from all objects extracted and find the most likely local lens properties these features constrain in a self-consistent manner. 
From this process, we find six features in all multiple images as shown in Fig.~\ref{fig:all_features} (left) and highlight them in different colours for visually easy matching. 
Their coordinates are listed in Table~\ref{tab:all_features} together with the uncertainties in their positions as inserted into our lens-model independent approach to constrain the local lens properties. 
\news{Fig.~\ref{fig:feature_comp} shows a comparison to the features listed in \cite{bib:Massey2018} (centre) and \cite{bib:Chen} (right).
While our best-fit features show coincidences, not all features of these previous works can be identified with more than 3-$\sigma$ significance by our pipeline.
Thus, a comparison of the local lens properties as constrained by all different feature sets also probes the impact of the different image-pre-processing pipelines on the mass-density reconstruction.}

\section{Optimum feature matching}
\label{sec:optimum_feature}

The choice as to which brightness features to match across multiple images greatly influences the lens reconstruction. 
In the previous works on A3827, a mere handful of choices were tested and only \cite{bib:Massey2015} set up two different feature matchings the authors deemed feasible\footnote{They already stated that the size of the offset depends on the feature identification and matching (see also Section~\ref{sec:sidm}).}. 
To follow-up, \cite{bib:Massey2018} collected complementary data of integral field spectroscopy. 
The CO(2-1)-transition observed with the Atacama Large Millimetre Array (ALMA), listed in Table~\ref{tab:filter_bands}, clearly showed that image~4 of \cite{bib:Carrasco} actually is two multiple images close together and that no third image of the central bulge appears in this region, either, as had been presumed in \cite{bib:Massey2015} and had been used in \cite{bib:Taylor} as well. 

Apart from this investigation, no further analyses were performed to alter the matching and track the impact of the changes on the lens reconstruction. 
However, since all lens reconstructions used these features to fit a lens model to the data, the impact can only be evaluated very indirectly. 
For instance, we can compare the quality-of-fit measures obtained after the fitting process, as detailed further in \cite{bib:Wagner_cluster}, or we compare the entailed conclusions about the offset between light and mass of the central galaxies (see Table~\ref{tab:related_work}).
In any case, for each combination of features, a full lens-modelling procedure is required, which is often computationally costly and time consuming. 

In contrast, reducing the evaluation of multiple-image configurations to their data-inferred local lens properties \citep{bib:Tessore, bib:Wagner_summary}, we could show that a shift in the position of a feature in one direction altered the corresponding reduced shear component inferred by our lens-model-independent local lens reconstruction for the much simpler Hamilton's Object in \cite{bib:Lin}. 
This insight implies that we can directly relate changes in the feature locations to changes in the local lens properties, which will be used for A3827 to reject improbable matchings in this section.  

\comm{After Section~\ref{sec:constraining_local} briefly introduces our method to constrain local lens properties without using a global mass density profile as a lens model,} Section~\ref{sec:joint_best-fit} describes the algorithm to find the joint optimum feature set and the corresponding best-fit local lens properties.
To study the impact of feature matchings on the local lens properties, Section~\ref{sec:impact_feature} shows a comparison between two different matchings, which serves as an example to motivate our procedure to jointly determine the optimum feature set and their local lens properties.
Then, based on our optimum features in Table~\ref{tab:all_features}, Section~\ref{sec:impact_number} investigates the persistence of local lens properties when changing the number of images used to infer the local lens properties.
Since the multiple images in A3827 have an extent that is \comm{of the order of their Einstein radius}, Section~\ref{sec:impact_location} analyses the change in local lens properties when changing the area covered by different feature combinations in each multiple image. 
\comm{In this way, the impact of higher-order lensing distortions can be studied for the first time, as \cite{bib:Wagner7} did not find any need to extend the lens-model-independent approach beyond leading order for observations known so far.}
The results are also discussed in Section~\ref{sec:comparative_discussion} in comparison to the ones obtained from the literature in Table~\ref{tab:related_work} to find interpretations of this unusual multiple-image configuration which bring all results into consistency with each other.

\subsection{Constraining local lens properties}
\label{sec:constraining_local}

A general overview of gravitational lensing in arbitrary spacetimes can be found in \cite{bib:Fleury2} or any standard text book on gravitational lenses, for instance, \cite{bib:SEF} or \cite{bib:Petters}.
Our approach employed here is based on the single-lens-plane formalism for strong gravitational lensing in a concordance $\Lambda$-Cold-Dark-Matter cosmology.

All two-dimensional positions in the lens plane are denoted by $\boldsymbol{x}$ and the angular diameter distance to the lens plane is $D_\mathrm{d}$ corresponding to its redshift $z_\mathrm{d}$. 
Analogously, we denote two-dimensional positions in the source plane by $\boldsymbol{y}$, its angular diameter distance is $D_\mathrm{s}$ corresponding to its redshift $z_\mathrm{s}$. 
The distance between lens and source plane is $D_\mathrm{ds}$.
Multiple images are indexed by $i,j = 1, 2, 3, ...$ and positions of brightness features by $\mu = 1, 2, 3, ...\,$. 

\begin{figure}
\centering
\includegraphics[width=0.45\textwidth]{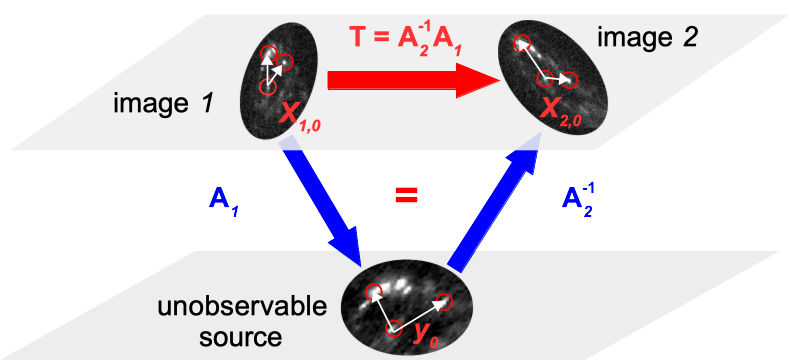}
\caption{Sketch of the \texttt{ptmatch} approach: corresponding brightness features (red circles) are selected in multiple images, so that the system of equations of Eq.~\ref{eq:matching} determines the local lens properties in Eqs.~\ref{eq:f}, \ref{eq:g}, and \ref{eq:j}. A linear transformation $\rm{T}$ between two images close to corresponding multiple-image positions $\boldsymbol{x}_{1,0}$, $\boldsymbol{x}_{2,0}$ can be determined from the brightness feature vectors (white arrows). This $\rm{T}$ corresponds to the product of the distortion matrices of these multiple images, enabling \texttt{ptmatch} to solve for the local lens properties.}
\label{fig:ptmatch}
\end{figure}

First, we summarise the approach of \cite{bib:Wagner2} and \cite{bib:Wagner_cluster} to match features in multiple images onto each other in order to extract local lens properties. 
The linearised lens equation around a point $\boldsymbol{x}_{i,0}$ in multiple image $i = 1, 2, 3, ...$ maps vectors around $\boldsymbol{x}_{i,0}$ into vectors around the source point $\boldsymbol{y}_{0}$ in the source plane.
There, $\boldsymbol{y}_{0}$ is the source position common to all multiple image points $\boldsymbol{x}_{i,0}$. 
Thus, with the distortion matrices ${\rm{A}}(\boldsymbol{x}_{i,0})$, we obtain
\begin{equation}
\boldsymbol{y}_{\mu} - \boldsymbol{y}_{0} = {\rm{A}}(\boldsymbol{x}_{i,0}) \left(\boldsymbol{x}_{i,\mu} - \boldsymbol{x}_{i,0} \right) = {\rm{A}}(\boldsymbol{x}_{j,0}) \left(\boldsymbol{x}_{j,\mu} - \boldsymbol{x}_{j,0} \right) \;,
\label{eq:matching}
\end{equation}
meaning that the lens mapping back-projects vectors between corresponding brightness features $\mu=1,2,...$ in multiple images (see Fig.~\ref{fig:A3827_CL0024} for two example configurations), here image~$i$ and image~$j$, onto the same vector in the source plane (see also Fig.~\ref{fig:ptmatch}). 

\comm{We assume that the $\boldsymbol{x}_{i,\mu}$ are so close to the $\boldsymbol{x}_{i,0}$, such that} the distortion matrix contains the convergence $\kappa$ and the shear components $\gamma_1$ and $\gamma_2$ at position $\boldsymbol{x}_{i,0}$.
The former is a measure of the local scaled mass density which scales the source properties by an overall factor to arrive at the size of the multiple image at position $\boldsymbol{x}_{i,0}$. 
The components of the shear represent the local leading-order distorting strength of the lens at $\boldsymbol{x}_{i,0}$.
\comm{Thus, higher-order distortions, like flexion, are neglected by construction keeping the area covered by all $\boldsymbol{x}_{i,\mu}$ small, as justified in \cite{bib:Wagner7}.}

Yet, neither convergence nor shear are directly constrained by the data and we have to rewrite Eq.~\ref{eq:matching} to obtain those local lens properties that the data actually constrain. 
First, we rewrite ${\rm{A}}_i \equiv {\rm{A}}(\boldsymbol{x}_{i,0})$ in terms of reduced shear components $g_1$ and $g_2$ as
\begin{align}
{\rm{A}}_i &= \left( \begin{matrix} 1 - \kappa_i - \gamma_{i,1} & -\gamma_{i,2} \\ -\gamma_{i,2} & 1 -\kappa_i + \gamma_{i,1} \end{matrix} \right) \nonumber \\
&= \left( 1 - \kappa_i \right) \left( \begin{matrix} 1 - g_{i,1} & -g_{i,2} \\ -g_{i,2} & 1 + g_{i,1} \end{matrix} \right) \,, \quad i=1, 2, ...\;.
\label{eq:A}
\end{align} 
Since the vectors in the source plane cannot be observed, we identify corresponding features in multiple images to set up the vectors $\left(\boldsymbol{x}_{i,\mu} - \boldsymbol{x}_{i,0} \right)$ for $i=1, 2, 3, ...$.
Then, we use the last part of Eq.~\ref{eq:matching} to write down a system of equations that links the vectors in different multiple images and their unknown distortion matrices.
While \comm{we can map all images onto each other, in principle}, the implementation we use, called \ptmatch\footnote{Available at \url{https://github.com/ntessore/imagemap}.}, fixes a so-called `reference image' with respect to which all \comm{local lens properties} are determined for the sake of efficiency. 
As in all related works, we set the reference image to be called image~$1$, which also applies to A3827. 
\comm{Solving the system of equations defined by the last part of Eq.~\ref{eq:matching} we obtain} the ratios of convergences between multiple images \comm{and reference image~1}
\begin{equation}
f_i = \dfrac{1-\kappa_1}{1-\kappa_i} \;, \quad i = 2, 3, ...
\label{eq:f}
\end{equation}
and the reduced shear components at all multiple image positions\footnote{It might seem counter-intuitive that the individual reduced shears can be reconstructed from products of distortions matrices. But one must keep in mind that, in single-plane lensing, distortion matrices are symmetric while their products are generally not. The antisymmetric part of those products contains information that contributes to the reconstruction of the individual (reduced) shears.}
\begin{equation}
g_{i,1} = \dfrac{\gamma_{i,1}}{1-\kappa_i} \;, \qquad g_{i,2} = \dfrac{\gamma_{i,2}}{1-\kappa_i} \;, \quad i=1, 2, 3, ...
\label{eq:g}
\end{equation}
with the amplitude and direction of the shear of image~$i$
\begin{equation}
	|\boldsymbol{g}|_i = \sqrt{g_{i,1}^2 + g_{i,2}^2}\;, \quad \varphi_i = \frac12 \text{atan} \left( \dfrac{g_{i,2}}{g_{i,1}} \right) \;,
	\label{eq:g_magnitude}
\end{equation}
if we have a minimum of three multiple images with two non-parallel vectors that can be matched in each of them. 
\comm{Figure}~\ref{fig:ptmatch} sketches the matching process, \cite{bib:Wagner2} details the requirements and degeneracies when solving the system of equations and \cite{bib:Wagner_cluster} describes the implementation. 
\comm{Additionally,} from Eqs.~\ref{eq:f} and \ref{eq:g} using Eq.~\ref{eq:A}, the magnification ratios
\begin{equation}
\mathcal{J}_i = \left(\det \left({\rm{A}}_i\right) \right)^{-1} \det \left({\rm{A}}_1\right) \;, \quad i=2,3, ...
\label{eq:j}
\end{equation}
can be calculated. 
For multiple images with a contiguous surface brightness profile or dominated by the central bulge like Hamilton's Object, observed flux ratios between pairs of images can be compared to these magnification ratios \citep{bib:Griffiths}. 
Differences between these magnification ratios and the flux ratios can hint at additional microlensing of individual images or dust extinction due to intervening gas clouds along the line of sight. 
Yet, since the fluxes for the highly detailed multiple images in A3827 are hard to define and measure, the magnification ratios are only useful to compare different lens reconstructions and approaches with each other. 

\comm{If the number of observables exceeds the minimum requirements, the system of equations defined by the last part of Eq.~\ref{eq:matching} can be solved for the ratios of convergences and reduced shear components, Eqs.~\ref{eq:f} and \ref{eq:g}, in a total-least-squares approach described in \cite{bib:Wagner2} and \cite{bib:Wagner_cluster}.
As also detailed in these papers, the confidence bounds on Eqs.~\ref{eq:f} and \ref{eq:g} are obtained by importance sampling from the full likelihood distribution of the $f$- and $\boldsymbol{g}$-values for the given observed positions, such that the local lens properties are the most likely and mean $f$- and $\boldsymbol{g}$-values with their 68\% confidence intervals, $\sigma$, corresponding to 1-$\sigma$ confidence bounds.
The overall quality of the fit is thus measured by the reduced $\chi^2$, $\chi_{\rm red}^2$, and the number of effective samples of the importance sampling, $n_{\rm s}$.}

Only ratios of the local $\kappa$, $\gamma_1$, $\gamma_2$ can be determined in this way because all observables are angular distances on the celestial sphere. 
\cite{bib:Wagner4} and \cite{bib:Wagner6} explain that the quantities in Eqs.~\ref{eq:f}, \ref{eq:g}, and \ref{eq:j} are not subject to any standard strong-lensing degeneracies and that they are indeed the maximum information retrievable from the observables without additional assumptions about the overall mass distribution.
To be converted to physical distances, they require an overall size scale, a physical distance, to be fixed. 
In standard gravitational lens reconstructions, the angular diameter distances are used which are based on a chosen cosmological model or on observations, like discussed in \cite{bib:Wagner5} to extend our approach to become independent of a specific parametrisation of a Friedmann cosmology.  

As \comm{will be} shown in Section~\ref{sec:impact_feature}, Eq.~\ref{eq:g_magnitude} can be employed to compare the local distorting properties of the lens to the visible matter distribution that could have caused this (reduced) shear. 
If there is a strong disagreement between the amplitude and direction of the inferred and the observationally motivated distortion, possible causes may be found in the limitations \comm{of} our approach like changing local lens properties over the area of a multiple image or a light-deflecting mass distribution that has a non-negligible extension along the line of sight. 

\begin{figure}
  \centering
  \includegraphics[width=0.4\textwidth]{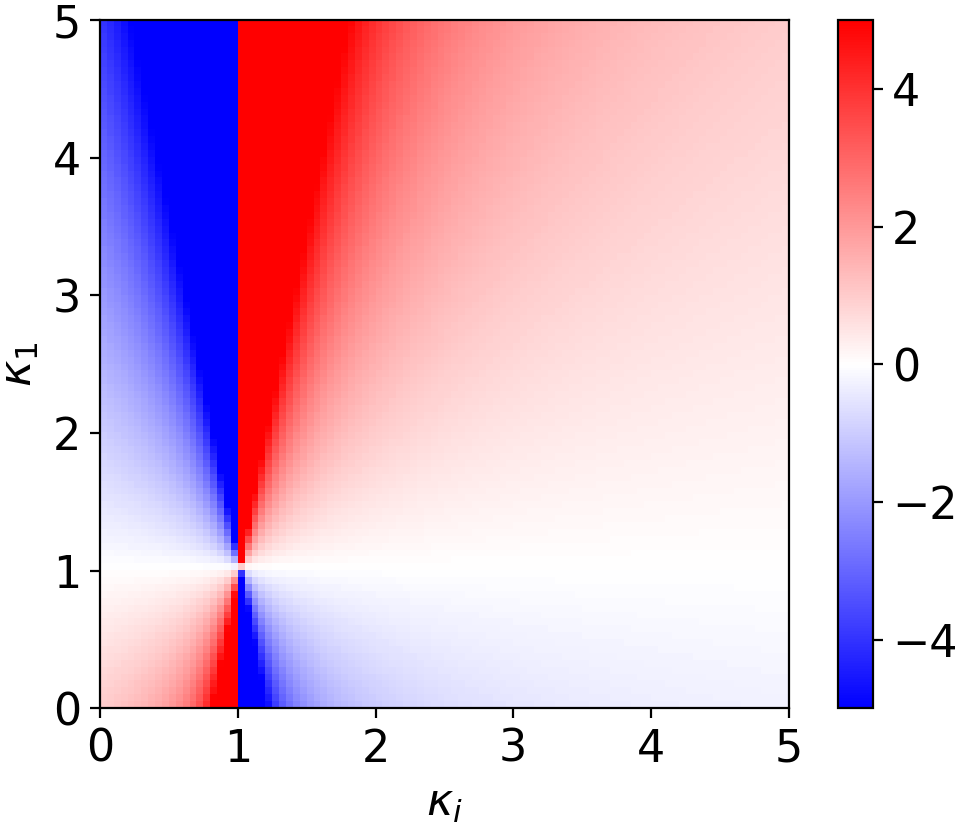}
  \caption{Resolving the degeneracy in the ratio of convergences: $f_i = (1-\kappa_1)/(1-\kappa_i)$ for varying $\kappa_1 \in \left[ 0, 5\right]$ and $\kappa_i \in \left[ 0, 5\right]$ (red colour for $f_i > 0$, blue for $f_i < 0$). Either fixing $\kappa_1 < 1$ or $\kappa_1 > 1$ and then deriving the relative $\kappa_i$ with respect to this choice has to coincide with the expected mass densities as inferred from complementary data like visible galaxies.}
  \label{fig:f_ratios}
\end{figure}

Similar to this, Eq.~\ref{eq:f} can be used to infer the relative convergences between the different images \comm{and compare the results to expectations from the visible light distribution and the kinematical status of the individual parts of the lens}. 
As Fig.~\ref{fig:f_ratios} details, there are two possible options, depending on $\kappa_1$ being smaller or larger than one. 
To decide which option is physically reasonable, one can use the rule of thumb that the mass density of a relaxed mass density profile is supposed to decrease from the centre outwards.
This assumption was already employed to conclude that all $\kappa_i$ in Hamilton's Object were smaller than one \citep{bib:Griffiths}. 

\begin{figure*}
\centering
\hspace{2ex} \includegraphics[width=0.4\textwidth]{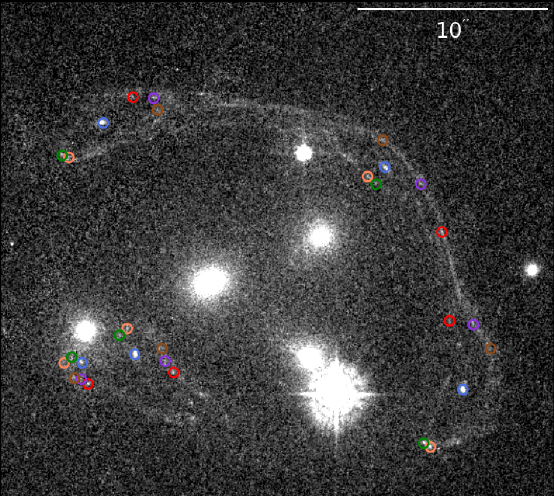} \hfill
\includegraphics[width=0.4\textwidth]{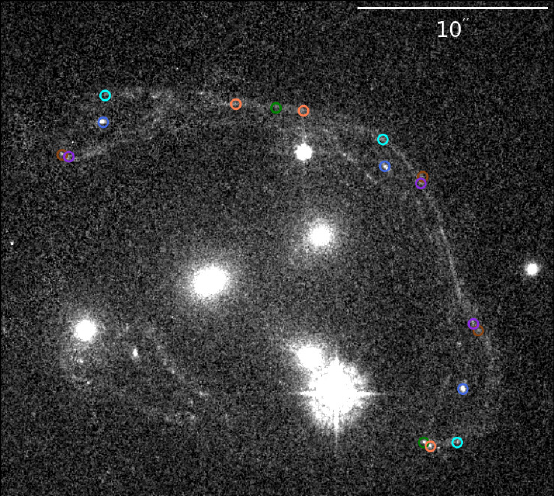} \hspace{2ex}
\caption{Six most persistent features with at least 3-$\sigma$ significance above local noise level (coloured circles) as identified by our persistence pipeline and the self-consistent matching in Section~\ref{sec:optimum_feature} (left). Alternative CL0024-like feature matching for images~1--3 with similarly robust features as expected from standard lensing theory but not favoured by the optimum feature matching procedure (right).
}
\label{fig:all_features}
\end{figure*}

\subsection{Joint optimum feature set and best-fit lens properties}
\label{sec:joint_best-fit}

As in previous works, we start to find robust corresponding features in images~1--3 and extend our process to images~4 and 5 only afterwards. 
But, we also support this process in Section~\ref{sec:impact_number} by showing that the local lens properties do not significantly depend on the choice of multiple images to be analysed.  

Based on all objects identified in the multiple images via the persistence pipeline of Section~\ref{sec:brightness_feature}, all possible combinations of features based on these objects could be inserted into \ptmatch and the quality-of-fit criteria, \comm{$\chi_{\rm red}$, $n_{\rm s}$, and the widths of the 1$\sigma$-confidence bounds of the $f$- and $\boldsymbol{g}$-values (see Section~\ref{sec:constraining_local})}, ranked. 
Since run-times of \ptmatch are of the order of fractions of a second, systematically ranking \comm{all possible combinations} is computationally feasible. 
We restrict our systematic testing to a subset of \comm{those combinations} that are physically plausible, meaning, a few hundred matchings that create similar feature matchings as observed in CL0024 (Fig.~\ref{fig:A3827_CL0024}, left) or a similar one to those of previous works on A3827. 
In this way, the quality-of-fit criteria (QFC) can still be evaluated alongside visual inspection of the individual feature matchings and a deeper understanding of the impact of relationship between feature positions and local lens properties can be gained. 

Fig.~\ref{fig:all_features} shows two possible feature matchings we considered, one inspired by the matching of previous works (left) and one inspired by the multiple-image configuration observed in CL0024 (right). 
From all matchings analogous to CL0024 that we analysed, the latter was found to be the optimum one and it turned out to have worse QFC than the matching similar to the ones in previous works. 
To define the ``optimum'' matching and find it among all possible sets, our systematic analyses of matchings motivates the following procedure to efficiently discard physically implausible or numerically unstable matchings and subsequently rank the remaining models to find the best one. 
\begin{enumerate}
\item Discard all matchings with $\chi^2_\mathrm{red} > 10$ because any $\chi^2_\mathrm{red} \gg 1$ implies that the degrees of freedom in \ptmatch do not suffice to adequately represent the input data and the local lens properties are biased. 
\item From the remaining set of possible matchings, discard all matchings with an effective number of samples $n_\mathrm{s} < 1000$.
Having effective numbers of samples of less than 10\% of the total number of samples used to set up the confidence bounds on the local lens properties hints at highly skewed probability distributions around the mean local lens property.
This, in turn, implies that the feature matching most likely leads to biased local lens properties. 
\item From the remaining set of possible matchings, discard all those matchings for which at least one local lens property has a \comm{confidence bound} $\sigma > 1$ and $\sigma > 3 \left| m \right|$.
In the last inequality, $m$ denotes the mean value of a local lens property. 
Sorting these matchings out, we discard numerically unstable solutions that may have their origin in the features spanning too small an area or being located too close to isocontours in the convergence with $\kappa=1$ (see \cite{bib:Wagner7} for further details). 
While these matchings may still be physically feasible, there is no reason to consider these solutions as optimum. 
Yet, if the set of optimum matchings after this process is empty, one may relax this step to find at least a physically plausible solution even if the features only constrain numerically unstable local lens properties. 
\item For the remaining set of possible matchings, set up four tables and rank all matchings within each table according to
\begin{itemize}
\item Table~A: minimum $\left| \chi^2_\mathrm{red} - 1 \right|$ with a preference for slight overfitting than for slight biasing,
\item Table~B: maximum number of features in a matching,
\item Table~C: minimum $\sigma_\mathrm{tot}=\sqrt{\sum \sigma^2 /n_\mathrm{i}}$, in which the sum is taken over all \comm{confidence bounds} $\sigma$ of all local lens properties for one matching and the result is divided by the number of multiple images $n_\mathrm{i}$ used in this matching, and
\item Table~D: maximum number of effective samples $n_\mathrm{s}$. 
\end{itemize}
In a similar manner as \grale has fitting criteria to evaluate the quality of fit for a lens model, these four tables provide a multi-dimensional QFC. 
Depending on the features available and the multiple-image configuration at hand, we are thus flexible to emphasise on different aspects of the quality when defining and finding the optimum feature matching and the corresponding best-fit local lens properties.
\end{enumerate}

\subsection{Impact of feature matching on local lens properties}
\label{sec:impact_feature}

Comparing the QFCs of Section~\ref{sec:joint_best-fit} for the two possible feature matchings shown in Fig.~\ref{fig:all_features} when restricting both results to the local lens properties of images~1--3, we find for the left one A: $\chi^2_\mathrm{red}=3.03$, B: six features, C: $\sigma_\mathrm{tot}=0.84$, D: $n_\mathrm{s}=3761$ (see also option~D in Table~\ref{tab:impact_number}) and for the right one A: $\chi^2_\mathrm{red}=9.97$, B: six features, C: $\sigma_\mathrm{tot}=1.20$, D: $n_\mathrm{s}=3462$ (see also option~C in Table~\ref{tab:impact_number}). 
To add a third matching to the four tables, we consider the matching of \cite{bib:Massey2018} restricted to images~1--3, such that the eight features named ``o,a,b,c,d,e,g,h'' can be inserted into \ptmatch\!.
For this matching, we obtain A: $\chi^2_\mathrm{red}=15.42$, B: eight features, C: $\sigma_\mathrm{tot}=0.09$, D: $n_\mathrm{s}=9937$ (see also option~A in Table~\ref{tab:impact_number}). 
Valuing Table~A more than any other table, the ranking clearly prefers the feature matchings obtained by our persistence pipeline over the manually identified ones, while focussing on the other criteria, the matching by \cite{bib:Massey2018} is favoured. 
Furthermore, since $\chi^2_\mathrm{red}$ for the CL0024-like matching is three times higher than the one of the anomalous matching and all other criteria are at least equal or even better for the latter, we define our optimum set of features to be the one shown in Fig.~\ref{fig:all_features} (left). 

\begin{figure}
\centering
\includegraphics[width=0.4\textwidth]{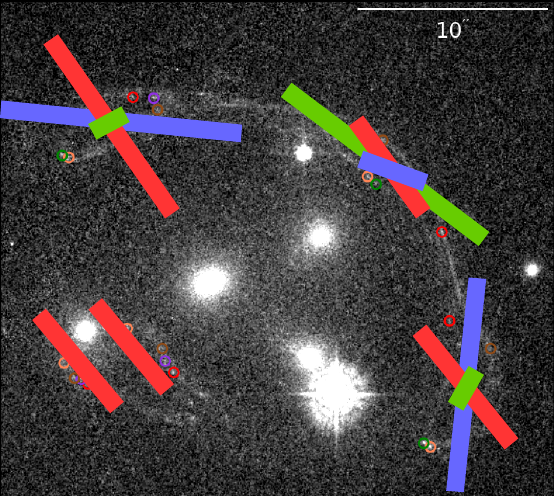} 
\caption{\textcolor{magenta}{Amplitudes and directions of reduced shear, obtained by \ptmatch for different choices of feature matchings: using our best-fit features of Fig.~\ref{fig:all_features}, left (red bars), the alternative CL0024-like matching of Fig.~\ref{fig:all_features}, right for images~1--3 (blue bars), and the matching of the features of \protect\cite{bib:Massey2018} for images~1--3 (green bars). 
Data to calculate \comm{$|\boldsymbol{g}|_i$ and $\varphi_i$} can be found in Table~\ref{tab:impact_number} option~B (green bars), option~C (blue bars), and option~G (red bars).}}
\label{fig:all_features_fit}
\end{figure}

A further point to consider is compatibility with standard lens theory and physical plausibility of the resulting local lens properties. 
For instance, we can use Eq.~\ref{eq:g} to obtain the amplitude and direction of the reduced shear for our best-fit features and alternative features from Fig.~\ref{fig:all_features} (left) and (right) from options~D, and C in Table~\ref{tab:impact_number}, respectively. 
We repeat the calculation for the \ptmatch results obtained for the three features of \cite{bib:Massey2018} that can be matched across images~1--5 (see option~B in Table~\ref{tab:impact_number}). 
The resulting reduced-shear amplitudes and directions are summarised in Fig.~\ref{fig:all_features_fit} as indicated by the lengths and directions of the coloured bars. 
The red ones represent the values retrieved from our best-fit features, the blue ones from our alternative CL0024-like features, and the green ones are obtained from the \cite{bib:Massey2018} features.
\news{As can be read off Table~\ref{tab:impact_number}, the reduced shear for images~4 and 5 of the \cite{bib:Massey2018} matching is subject to broad confidence bounds. 
No clear amplitude or direction can be determined. 
Therefore, Fig.~\ref{fig:all_features_fit} does not show them.}

From Fig.~\ref{fig:all_features_fit}, we see that the matching is directly related to the inferred local lens properties, such that ambiguities in the matching have a strong impact on the (local) lens reconstruction. 
The matching of points in images~4 and 5, for instance, defines whether the two images form a standard fold like images~4 and 5 in CL0024 (our optimum feature matching, red arrows) or whether the two images have the same parity (feature matchings by \cite{bib:Chen} and \cite{bib:Massey2018}, \news{both of which have confidence bounds exceeding the calculated values}).
Such a fold configuration consisting of images~4 and 5, having absolute reduced shear amplitudes close to one and having almost identical directions has also been found for Hamilton's Object \citep{bib:Griffiths} and is thus in agreement with standard lensing theory and physically plausible given that there is a luminous mass, G1, to which this shear can be attributed. 
The preference for image~4 \news{with $|\boldsymbol{g}_4| \in \left[ 0.96, 1.09\right]$ within 1-$\sigma$ confidence} to have $\left| \boldsymbol{g} \right] \ge 1$ and image~5 \news{with $|\boldsymbol{g}_5| \in \left[ 0.91, 0.98\right]$ within 1-$\sigma$ confidence} to have $\left| \boldsymbol{g} \right] \le 1$ is in contradiction with the expected values which should just show the opposite behaviour. 
Nevertheless, the fact that image~4 is closer to G1 could explain this anomaly. 

In contrast, \cite{bib:Massey2018} chose a feature matching for images~4 and 5 such that both images have the same parity and thus do not form a standard fold configuration. 
This same-parity matching directly propagates through to the parities of the local lens properties. 
As can be seen when comparing the signs of the magnification ratios for images~4 and 5, the feature matching of \cite{bib:Massey2018} yields the same sign, see option~B in Table~\ref{tab:impact_number}.
In the same way, the magnification ratios of these images for our best-fit features, option~G in Table~\ref{tab:impact_number}, consistently have different signs. 
Thus, the feature matching directly determines the relative parity between multiple images, as also noted in earlier works \citep{bib:Wagner_cluster, bib:Griffiths, bib:Lin}.

For the same-parity configuration of images~4 and 5, the amplitudes and directions of the reduced shear are not those typical of a standard fold configuration.
While this choice of equal parities in the matching of \cite{bib:Massey2018} is unusual, it may be possible to explain it in terms of a higher-order singularity than the standard fold, as, for instance outlined for other cases in \cite{bib:Orban} and theoretically explained in \cite{bib:Petters}.
\comm{Figure}~\ref{fig:all_features_fit} only shows the matching of \cite{bib:Massey2018}, but from option~C in Table~\ref{tab:comparison_reconstructions}, we read off a similar result for the matching of \cite{bib:Chen}. 

\begin{table*} 
\caption{Local lens properties as obtained from \ptmatch and their dependency on the images used: 
A: using features ``o,a,b,c,d,e,g,h'' from \protect\cite{bib:Massey2018} from images~1--3; 
B: using features ``o,a,c'' from \protect\cite{bib:Massey2018} from images~1--5 (see also option~B of Table~\ref{tab:comparison_reconstructions}; 
C: using our alternative CL0024-like features from images~1--3 (see Fig.~\ref{fig:all_features}, right); 
D: same as C but for our best-fit features (see Fig.~\ref{fig:all_features}, left); 
E: same as D but for images~1--4;
F: same as D but with image~5; 
G: same as D but for images~1--5 (see also option~A of Table~\ref{tab:comparison_reconstructions}).
\vspace{-1.5ex}
}
\label{tab:impact_number}		
\begin{center}
\begin{tabular}{c|rr|rr|rr|rr|rr|rr|rr}
\hline
  & \multicolumn{2}{c|}{A} & \multicolumn{2}{c|}{B} &  \multicolumn{2}{c|}{C} &  \multicolumn{2}{c|}{D} &  \multicolumn{2}{c|}{E} &  \multicolumn{2}{c|}{F} &  \multicolumn{2}{c|}{G}   \\
\hline
  & \multicolumn{2}{c|}{$\chi^{2}_{\rm red}$ = 15.42} & \multicolumn{2}{c|}{$\chi^{2}_{\rm red}$ = 0.23} & \multicolumn{2}{c|}{$\chi^{2}_{\rm red}$ = 9.97} & \multicolumn{2}{c|}{$\chi^{2}_{\rm red}$ = 3.03} & \multicolumn{2}{c|}{$\chi^{2}_{\rm red}$ = 3.36} & \multicolumn{2}{c|}{$\chi^{2}_{\rm red}$ = 4.09} & \multicolumn{2}{c|}{$\chi^{2}_{\rm red}$ = 4.09}  \\
    & \multicolumn{2}{c|}{$n_{\rm f} = 8$} & \multicolumn{2}{c|}{$n_{\rm f} = 3$} & \multicolumn{2}{c|}{$n_{\rm f} = 6$} & \multicolumn{2}{c|}{$n_{\rm f} = 6$} & \multicolumn{2}{c|}{$n_{\rm f} = 6$} & \multicolumn{2}{c|}{$n_{\rm f}=6$} & \multicolumn{2}{c|}{$n_{\rm f} = 6$}\\
    & \multicolumn{2}{c|}{$\sigma_{\rm tot} = 0.09$} & \multicolumn{2}{c|}{$\sigma_{\rm tot} = 279.73$} & \multicolumn{2}{c|}{$\sigma_{\rm tot} = 1.20$} & \multicolumn{2}{c|}{$\sigma_{\rm tot} = 0.84$} & \multicolumn{2}{c|}{$\sigma_{\rm tot} = 4.07$} & \multicolumn{2}{c|}{$\sigma_{\rm tot} = 0.33$} & \multicolumn{2}{c|}{$\sigma_{\rm tot} = 0.37$}\\
  & \multicolumn{2}{c|}{$n_{\rm s} = 9937$} &  \multicolumn{2}{c|}{$n_{\rm s} = 3763$} & \multicolumn{2}{c|}{$n_{\rm s} = 3462$} & \multicolumn{2}{c|}{$n_{\rm s} = 3761$} & \multicolumn{2}{c|}{$n_{\rm s} = 3249$} & \multicolumn{2}{c|}{$n_{\rm s} = 6055$} & \multicolumn{2}{c|}{$n_{\rm s} = 5662$}  \\
  \hline
  & Mean & Std & Mean & Std & Mean & Std & Mean & Std & Mean & Std & Mean & Std & Mean & Std  \\
\hline
$g_{1,1}$ & 0.32 & 0.02 & -0.18 & 0.06 & -1.79 & 0.20 & -0.27 & 0.05 & -0.25 & 0.05 & -0.30 & 0.04 & -0.29 & 0.04  \\
$g_{1,2}$ & -0.20 & 0.02 & 0.30 & 0.11 & 0.35 & 0.05 & -1.36 & 0.26 & -1.51 & 0.28 & -1.13 & 0.18 & -1.21 & 0.19  \\
\hline
$\mathcal{J}_{2}$ & -0.39 & 0.03 & -0.47 & 0.12 & -0.78 & 0.09 & -0.61 & 0.04 & -0.61 & 0.04 & -0.62 & 0.04 & -0.62 & 0.04 \\
$f_{2}$ & 0.65 & 0.07 & 1.40 & 1.30 & 0.48 & 0.11 & 0.25 & 0.04 & 0.24 & 0.03 & 0.29 & 0.03 & 0.27 & 0.03 \\
$g_{2,1}$ & 0.59 & 0.02 & 0.59 & 0.46 & 0.44 & 0.06 & -0.31 & 0.04 & -0.31 & 0.04 & -0.31 & 0.04 & -0.31 & 0.04  \\
$g_{2,2}$ & 1.25 & 0.06 & -2.02 & 1.20 & -0.39 & 0.04 & -0.90 & 0.03 & -0.89 & 0.03 & -0.93 & 0.03 & -0.92 & 0.03 \\
\hline
$\mathcal{J}_{3}$ & 0.62 & 0.03 & 0.77 & 0.14 & 2.71 & 0.12 & 0.38 & 0.04 & 0.38 & 0.04 & 0.38 & 0.04 & 0.38 & 0.04  \\
$f_{3}$ & 0.46 & 0.02 & 0.88 & 0.10 & 2.46 & 1.86 & 1.31 & 0.36 & 1.54 & 2.22 & 1.09 & 0.16 & 1.16 & 0.21 \\
$g_{3,1}$ & 0.80 & 0.01 & 0.18 & 0.10 & 2.31 & 0.85 & -0.94 & 0.60 & -1.34 & 3.40 & -0.54 & 0.28 & -0.66 & 0.35  \\
$g_{3,2}$ & -0.28 & 0.03 & 0.26 & 0.14 & -0.41 & 0.15 & -2.27 & 1.24 & -3.12 & 7.05 & -1.41 & 0.54 & -1.67 & 0.68   \\
\hline
$\mathcal{J}_{4}$ & - & - & -0.25 & 0.07 & - & - & - & - & 0.01 & 0.01 & - & - & 0.01 & 0.01 \\
$f_{4}$ & - & - & 1.39 & 171.91 & - & - & - & - & -0.04 & 0.02 & - & - & -0.03 & 0.02 \\
$g_{4,1}$ & - & - &1.22 & 114.49 & - & - & - & - & -0.05 & 0.14 & - &  - & -0.18 & 0.10 \\
$g_{4,2}$ & - & - & -2.69 & 261.05 & - & - & - & - & -1.07 & 0.08 & - &  - & -1.01 & 0.05 \\
\hline
$\mathcal{J}_{5}$ & - & - & -0.20 & 0.06 & - & - & - & - & - & - & -0.09 & 0.02 & -0.09 & 0.02 \\
$f_{5}$ & - & - & 2.29 & 241.46 & - & - & - & - & - & - & 0.15 & 0.04 & 0.13 & 0.03  \\
$g_{5,1}$ & - & - & -4.58 & 438.96 & - & - & - & - & - & - & -0.22 & 0.09 & -0.18 & 0.09  \\
$g_{5,2}$ & - & - & -0.88 & 171.57 & - & - & - & - & - & - & -0.94 & 0.02 & -0.93 & 0.02  \\
\hline                                                          
\end{tabular}
\end{center}
\end{table*}

Comparing the ratios of convergences with those inferred in CL0024 or from Hamilton's Object and ordering the convergences according to their relative values as detailed in Fig.~\ref{fig:f_ratios}, we find that our best-fit features favour a solution for which $\kappa_5 > \kappa_2 > \kappa_1 > \kappa_3 > 1 >  \kappa_4$. 
The same result can be obtained from our alternative CL0024-like feature matching.
This order is physically plausible from the point of view that images~5 and 2 are closest to the cluster centre and should therefore be exposed to the largest mass density. 
However, it remains puzzling that image~4 is supposed to have a small $\kappa$ despite its close proximity to the image with the largest convergence. 
An edge in the mass density distribution, for instance, treating G1 as a perturbation to the smooth overall mass density, would have to be assumed.
\news{For the feature matchings of \cite{bib:Massey2018} and \cite{bib:Chen}, assuming $\kappa_1 > 1$ leads to a relative order of convergences of $\kappa_3 > \kappa_1 > \kappa_2 > 1$, which seems physically less plausible based on the expectation that $\kappa_2$ should be largest in the cusp configuration due to its closest proximity to the centre. 
Assuming $\kappa_1 < 1$, we obtain $1 > \kappa_2 > \kappa_1 > \kappa_3$, which agrees to the order expected from other cases and also to the order of our feature matchings.
}

Comparing the QFC as listed in Table~\ref{tab:impact_number} for these options, we note that the feature matching of \cite{bib:Massey2018} would be sorted out of the possible matchings because all local lens properties of images~4 and 5 except the magnification ratios violate the rule that $\sigma >1$ should have $\sigma \le 3 \left|m\right|$. 
\news{As we already saw in comparison to lens models in CL0024, the increasing confidence bounds can be explained by the images lying closely to the isocontour $\kappa=1$.
A comparison to the \lenstool and \grale reconstructions of \cite{bib:Massey2018} (see their Fig.~4) corroborates this explanation for A3827 as well.
Table~\ref{tab:comparison_reconstructions} shows that similarly broad confidence bounds for images~4 and 5 can be obtained from the matching of \cite{bib:Chen}. 
However, there is no convergence map in their paper to cross-check.}

\vspace{2ex}
Restricting the matching to images~1--3 for the \cite{bib:Massey2018} features \news{and thereby increasing the number of features}, leads to a high $\chi^2_\mathrm{red}$, such that the \cite{bib:Massey2018} matching is still ranked below the one of our best-fit features. 
Consequently, despite the remaining unresolvable degeneracy between the feature matchings, all these arguments favour a standard fold configuration for images~4 and 5, which has now been found for the first time when comparing local lens properties inferred from different possible matchings with each other. 

We note that even a matching inspired by the one in CL0024 or Hamilton's Object does not yield a relative reduced shear configuration pointing at a standard cusp for images~1--3, as could be derived for Hamilton's Object in \citep{bib:Griffiths}. 
Besides this, we could not find a fully convincing order of relative convergences.

Taken all findings together, our initial assumption is supported that the non-standard configuration of central galaxies has a strong impact on the multiple-image configuration being anomalous.
The feature matching directly influences the relative image parities and controls the local mass-density reconstructions.
\news{Ranking all reconstructions in four tables according to QFCs thus yields a list of best-fit reconstructions sorted by mathematical fitness.
However, a follow-up inspection of the physical plausibility as done here is still necessary.
As A3827 reveals, general QFC are hard to define and, depending on their definition, different solutions may be favoured.}

\subsection{Impact of number of images on local lens properties}
\label{sec:impact_number}

\begin{table*} 
\caption{Local lens properties as obtained from \ptmatch and their dependency on the number of features used:
A: using all our best-fit features (see Fig.~\ref{fig:all_features}, left); 
B: using only features~2--6; 
C: using only features~1--3;
D: using only features~1, 2, 6; 
E: using only features~1, 4, 6; 
F: using only features~1, 5, 6.
}
\label{tab:impact_location}
\begin{center}
\vspace{-1.5ex}
\begin{tabular}{c|rr|rr|rr|rr|rr|rr}
\hline
   & \multicolumn{2}{c|}{A} & \multicolumn{2}{c|}{B} &  \multicolumn{2}{c|}{C} &  \multicolumn{2}{c|}{D} &  \multicolumn{2}{c|}{E} &  \multicolumn{2}{c|}{F} \\
\hline
   & \multicolumn{2}{c|}{$\chi^{2}_{\rm red}$ = 4.09} & \multicolumn{2}{c|}{$\chi^{2}_{\rm red}$ = 3.31} & \multicolumn{2}{c|}{$\chi^{2}_{\rm red}$ = 8.44} &  \multicolumn{2}{c|}{$\chi^{2}_{\rm red}$ = 8.95} &  \multicolumn{2}{c|}{$\chi^{2}_{\rm red}$ = 0.22} & \multicolumn{2}{c|}{$\chi^{2}_{\rm red}$ = 5.01}  \\
    & \multicolumn{2}{c|}{$n_{\rm f} = 6$} & \multicolumn{2}{c|}{$n_{\rm f} = 5$} & \multicolumn{2}{c|}{$n_{\rm f} = 3$} & \multicolumn{2}{c|}{$n_{\rm f} = 3$} & \multicolumn{2}{c|}{$n_{\rm f} = 3$} & \multicolumn{2}{c|}{$n_{\rm f} = 3$} \\
    & \multicolumn{2}{c|}{$\sigma_{\rm tot} = 0.37$} & \multicolumn{2}{c|}{$\sigma_{\rm tot} = 0.24$} & \multicolumn{2}{c|}{$\sigma_{\rm tot} =  30.42$} &\multicolumn{2}{c|}{$\sigma_{\rm tot} = 14.25$}  & \multicolumn{2}{c|}{$\sigma_{\rm tot} = 2.90$} & \multicolumn{2}{c|}{$\sigma_{\rm tot} = 2.87$} \\
   & \multicolumn{2}{c|}{$n_\mathrm{s}$ = 5662} & \multicolumn{2}{c|}{$n_\mathrm{s}$ = 5882} & \multicolumn{2}{c|}{$n_\mathrm{s}$ = 128} &  \multicolumn{2}{c|}{$n_\mathrm{s}$ = 225}  &  \multicolumn{2}{c|}{$n_\mathrm{s}$ = 8413}  & \multicolumn{2}{c|}{$n_\mathrm{s}$ = 6618} \\
\hline
   & Mean & Std & Mean & Std & Mean & Std & Mean & Std & Mean & Std & Mean & Std  \\
\hline
$g_{1,1}$ & -0.29 & 0.04 & -0.30 & 0.04 & 0.25 & 0.29 & -0.59 & 0.03 & -0.26 & 0.04 & -0.22 & 0.07  \\
$g_{1,2}$ & -1.21 & 0.19 & -1.11 & 0.17 & 0.45 & 0.18  & 0.80 & 0.02 & -1.28 & 0.14 & -0.92 & 0.19  \\
\hline
$\mathcal{J}_{2}$ & -0.62 & 0.04 & -0.62 & 0.04 & -0.20 & 0.13 & -0.14 &0.17 & -0.57 & 0.18 & -0.69 & 0.18  \\
$f_{2}$ & 0.27 & 0.03 & 0.29 & 0.03 & 1.61 & 19.22 & -0.81 & 29.09 & 0.25 & 0.08 & 0.36 & 0.10  \\
$g_{2,1}$ & -0.31 & 0.04 & -0.32 & 0.04 & 2.12 & 25.32 & -0.57 & 9.83 & -0.33 & 0.04 & -0.20 & 0.08  \\
$g_{2,2}$ & -0.92 & 0.03 & -0.93 & 0.03 & -2.14 & 29.94 & 0.87 & 8.23 & -0.90 & 0.03 & -0.99 & 0.04 \\
\hline
$\mathcal{J}_{3}$ & 0.38 & 0.04 & 0.41 & 0.05 & 0.49 & 0.20 & 4.21 & 1.12 & 0.41 & 0.07 & 0.45 & 0.12  \\
$f_{3}$ & 1.16 & 0.21 & 1.06 & 0.12 & 0.54 & 0.18 & 6.14 & 1.11 & 1.86 & 2.69 & 1.06 & 0.45  \\
$g_{3,1}$ & -0.66 & 0.35 & -0.50 & 0.23 & 0.45 & 0.23 & -0.56 & 0.18 & -1.42 & 3.65 & -0.17 & 0.51  \\
$g_{3,2}$ & -1.67 & 0.68 & -1.32 & 0.42 & 0.54 & 0.24 & 0.77 & 0.16 & -2.39 & 4.61 & -0.90 & 0.83  \\
\hline
$\mathcal{J}_{4}$ & 0.01 & 0.01 & 0.01 & 0.01 & -0.03 & 0.06 & 0.65 & 0.35 & 0.10 & 0.03 & 0.04 & 0.03 \\
$f_{4}$ & -0.03 & 0.02 & -0.02 & 0.02 & 0.12 & 4.63 & 1.56 & 0.51 & -0.37 & 0.12 & -0.27 & 0.94  \\
$g_{4,1}$ & -0.18 & 0.10 & -0.23 & 0.09 & 0.85 & 3.41 & -0.61 & 0.18 & -0.20 & 0.14 & -0.18 & 2.41  \\
$g_{4,2}$ & -1.01 & 0.05 & -0.99 & 0.04 & -0.54 & 9.96 & 0.65 & 0.35 & -1.40 & 0.34 & -0.99 & 5.72 \\
\hline
$\mathcal{J}_{5}$ & -0.09 & 0.02 & -0.10 & 0.02 & -0.45 & 0.19 & -2.92 & 0.61 & -0.30 & 0.06 & -0.18 & 0.06 \\
$f_{5}$ & 0.13 & 0.03 & 0.15 & 0.03 & -2.12 & 28.49 & 4.43 & 0.92 & 0.40 & 0.08 & 0.55 & 0.27  \\
$g_{5,1}$ & -0.18 & 0.09 & -0.23 & 0.08 & 2.68 & 41.38 & -0.70 & 0.31 & -0.12 & 0.07 & -0.34 & 0.38 \\
$g_{5,2}$ & -0.93 & 0.02 & -0.94 & 0.02 & 0.57 & 7.87 & 0.75 & 0.08 & -0.78 & 0.08 & -1.11 & 0.40 \\
\hline                                                          
\end{tabular}
\end{center}
\end{table*}

Next, we investigate the impact of different images on the inferred local lens properties. 
We already found stability of the local lens properties over different filter bands in \cite{bib:Lin}, but, so far, we have not analysed whether adding or removing images from the analysis changes the local lens properties. 
As can read off options~D--G in Table~\ref{tab:impact_number}, we also observe stability for the local lens properties in that respect. 
This result is expected from the fundamental theoretical perspective that the local lens properties are directly constrained by the transformations of multiple-images onto each other which are calculated from the observables at these images \citep{bib:Petters, bib:Tessore, bib:Wagner_summary}. 
Hence, up to an overall not constrainable scale to fix the ratio of source to image size and a global distortion as detailed in \cite{bib:Wagner7}, the local lens properties for each image are determined by the mass density encountered by a light bundle along the path leading to this image.
They are thus decoupled as much as possible and, in particular, free from a global model assumption of a mass density profile in the lens plane.  
 
For the \cite{bib:Massey2018} features, we note that the local lens properties are less stable than for our best-fit features obtained with a robust image-processing pipeline.
50\% of the local lens properties for images~1--3 inferred from \cite{bib:Massey2018} features do not coincide within their 1-$\sigma$ confidence bounds when adding images~4 and 5 to the \ptmatch analysis (compare options~A and B in Table~\ref{tab:impact_number}), while all local lens properties inferred from our best-fit features do so (compare options~D--G). 
These changes can be attributed to the change in the number of features used from eight features in option~A to only three features in option~B because the latter also cover a smaller area of each multiple image (see Section~\ref{sec:impact_location} for more details).
\news{A cross-check running \ptmatch for the same three features as used in option~B for images~1--3 shows a coincidence of all local lens properties of image~1 and 3 within their 1-$\sigma$ confidence bounds. 
Yet, the feature matching for image~2 seems to be located too close to a $\kappa=1$ isocontour again as the confidence bounds exceed the calculated properties by far.
We can thus conclude that the number of images does not have an influence on the local lens properties irrespective of the feature set.}

%
%
%

\subsection{Impact of location of features on local lens properties}
\label{sec:impact_location}

Next, as also done for CL0024 in great detail in \cite{bib:Wagner_cluster}, we repeat the analysis of selecting different features of the six best-fit ones of Fig.~\ref{fig:all_features} (left) and track the changes in the local lens properties. 
Contrary to the results for CL0024, we expect the local lens properties to change because their extent is a significant portion of their Einstein radius and, as a consequence, the prerequisite for \ptmatch that the local lens properties remain constant over the area covered by the multiple images is violated. 
\comm{Nevertheless, restricting the \ptmatch analysis to smaller parts of the images, the prerequisite may still hold, such that we can gain an estimate for the length scale over which the light-deflecting mass density remains constant.}

Before we turn to this issue, we investigate the impact of the large extent of feature~1, already mentioned in Section~\ref{sec:feature_extraction} (see Fig.~\ref{fig:image245_objects}, left). 
For our best-fit feature set of Section~\ref{sec:summary_persistent}, all locations along the red, dotted line in Fig.~\ref{fig:image245_objects} (left) were used to find the optimum position of feature~1 in image~2 and the local lens properties summarised in option~A of Table~\ref{tab:impact_location}. 
Leaving this feature out and running \ptmatch on features~2--6 only, still, all QFC are improved, as can be read off Table~\ref{tab:impact_location} comparing option~A with B. 
Hence, doubling the measurement uncertainty for the position of feature~1, as mentioned in Table~\ref{tab:all_features} is justified by this result. 

\news{Reducing the features to run \ptmatch on a combination of features, such that different and smaller areas are covered for each image, we can investigate the constancy of the local lens properties.
We choose a combination of features~1--3 (option~C in Table~\ref{tab:impact_location}), features~1, 2, 6 (option~D in Table~\ref{tab:impact_location}), features~1, 4, 6 (option~E in Table~\ref{tab:impact_location}), and another one of features~1, 5, 6 (option~F in Table~\ref{tab:impact_location}).
Firstly, the $\chi_\mathrm{red}^2$-values hint at stronger biasing for options~ C, D, and F compared to our fiducial option~A. 
A second look reveals that they would be excluded from the list of feasible reconstructions according to the QFC set up in Section~\ref{sec:joint_best-fit}. 
Here, we explore them to understand the relation between the QFC and the physical properties of the reconstructions, but we exclude those multiple images from our analysis that violate the QFC set up in Section~\ref{sec:joint_best-fit}.} 

\news{Comparing the local lens properties of options~C--F with our fiducial option~A, it becomes clear that the latter is dominated by the area of the multiple images spanned by features~1-4, 5, and 6. 
For option~E, we find that \ptmatch is overfitting and thus, the local lens properties as determined from this feature combination can be considered in accordance with constant local lens properties again, like in previous examples. 
Only image~3 suffers from a numerical instability, which drives $\sigma_\mathrm{tot}$ to higher values.
Thus, we can now understand that option~A is biased with a $\chi_\mathrm{red}^2 > 1$ due to the changing local lens properties in the patch spanned by the features~1, 2, 3, and 6. 
Yet, its $\sigma_\mathrm{tot}$ shows lower values due to the increased area covered by all features, such that numerical stability is guaranteed from this point of view. 
Whether or not the numerical instabilities in options~C--F arise due to the small area covered by the features or whether the specific area is close to a $\kappa=1$ isocontour remains unknown.
However, based on the location of the patch, one explanation or the other may be more likely. 
The instabilies of option~C are most likely caused by the patch covering too small an area in each multiple images, as also supported by the analyses done in \cite{bib:Wagner_cluster}.}

\news{We can thus track the change of the local lens properties as inferred from option~C to those inferred from option~D, which covers a patch in the multiple image directly neighbouring the one from option~C. 
In the same way, we can transit from the patch used in option~D to those used in options~E and F.
Since the local lens properties of options~E are considered to be stable, this means that the distorting properties of the lens in the region close to G1 are rapidly changing. 
Therefore, any extrapolation of the local lens reconstruction outside the patch of option~E is already biased and statements on the potential offset between light and mass in G1 seem difficult from the lens-model-independent perspective (more details are discussed in Section~\ref{sec:offset_analysis}).}


%
%
%
%


\vspace{-2ex}
\section{Comparative discussion}
\label{sec:comparative_discussion}

Based on the results gained in Sections~\ref{sec:brightness_feature} and \ref{sec:optimum_feature}, and in the works listed in Table~\ref{tab:related_work}, we now comment on the offset between light and mass for G1 from the lens-model-independent point of view.
\comm{To do so, we investigate how fast the local lens properties inferred from the observables, as detailed in Section~\ref{sec:impact_location}, change and thereby, how strongly the mass density reconstruction at the position of the lensing galaxies depends on the lens-model assumptions or can still be estimated based on the data.} 
These evaluations are all contained in Section~\ref{sec:offset_analysis}. 

\news{Then, Section~\ref{sec:comparison_previous} compares our results to the related works in Table~\ref{tab:related_work}.
It particularly investigates the interpretations for the configuration of images~4 and 5 and discusses a potentially new kind of degeneracy found in this cluster. 
Based on strong-lensing data alone, it does not seem possible to distinguish between a single-lens-plane mass-density reconstruction with external shear and a multiple-lens-plane mass-density reconstruction with additional masses along the line of sight due to the mathematical degeneracy that a sequence of shearing and scalings can also be written as an effective shearing plus a rotation.}
Section~\ref{sec:conclusion} details the implications for dark-matter-candidate searches based on strong-lensing phenomena and gives guidelines to improve upon the strong-lensing data evaluation standards, also including complementary data sets.

In Section~\ref{sec:summary_contributions}, we summarise our contributions to gain a deeper understanding of the multiple-image configuration in A3827. 
It focusses on the strong-lensing effect and puts A3827 in the context of the two other multiple-image configurations we have analysed similarly, CL0024 and Hamilton's Object \citep{bib:Wagner_cluster, bib:Griffiths, bib:Lin}.
Moreover, it highlights the knowledge gained for A3827 in the least model-dependent way, supported by a robust feature extraction and a systematic process to identify the best matching.

\subsection{Offset analysis from a lens-model-independent viewpoint}
\label{sec:offset_analysis}

\news{As already detailed in \cite{bib:Wagner_summary} based on \cite{bib:Tessore}, the local lens properties are the maximum information about a strong gravitational that can be gained without making any assumptions about the global mass density profile. 
In \cite{bib:Wagner7}, this statement was refined based on the finding that the feature mapping between the multiple images only measures \emph{relative} local lens properties, such that an overall scaling and common distortion of all multiple images cannot be constrained by the observables.
Yet, these degeneracies do not affect any of the conclusions for the offset made from a lens-model-independent viewpoint. 
}

\news{In Section~\ref{sec:impact_location}, we found that the local lens properties inferred from the features~1, 4, 6 are the most stable ones.
They change rapidly for the other side of each multiple image mainly spanned by features~2 and 3, such that extrapolating outside the area covered by features~1, 4, and 6 is already subject to a bias, indicated by $\chi_\mathrm{red}^2$.
Thus, the analyses as performed in \cite{bib:Wagner1} to determine the region in which the local lens properties close to a standard fold configuration can be extrapolated cannot be applied here.
As a consequence, we cannot infer the lens properties at G1 or in its vicinity without making assumptions about the rate of change of the local lens properties and the smoothness of the lensing potential or the mass density.
This, in turn, implies that any statement of a potential offset is driven by the lens model used to reconstruct the mass density. 
Further support for this statement can be obtained by noting that the feature matchings of \cite{bib:Massey2018} and \cite{bib:Chen} are similar and their local lens properties also show a high degree of coincidence (comparing option~B and C in Table~\ref{tab:comparison_reconstructions}. 
But both works, based on different models and different evaluation approaches, arrive at different conclusions whether light traces mass in A3827.}

\news{While our best-fit features favour a standard fold configuration for images~4 and 5, changing this arrangement to a non-standard fold with two equal-parity images, as suggested in \cite{bib:Massey2018} and \cite{bib:Chen}, it is clear that the local lens properties take on different values compared to our fiducial reconstruction. 
As the equal-parity configuration of image~4 and 5 leads to a higher-order singularity which is less stable than a standard fold, its nature implies that the local lens properties are less robust than those for a standard fold. 
Consequently, we do not expect this feature matching to yield a more robust extrapolation of the local lens properties towards G1 to make a significant statement on the offset between light and mass.}

\begin{table} 
\caption{Local lens properties for images~1--5 of A3827 using different features sets inserted into \ptmatch\!. 
A: six features extracted by our persistence pipeline (same as option~G in Table~\ref{tab:impact_number}); 
B: use three features common to all images in \protect\cite{bib:Massey2018} (called ``o, a, c'') ; 
C: same as B but for the three common features of the matching in \protect\cite{bib:Chen} (called 1,2,4).
}
\label{tab:comparison_reconstructions}
\begin{center}
\vspace{-2ex}
\begin{tabular}{c|rr|rr|rr}
\hline
\noalign{\smallskip}
  & \multicolumn{2}{c|}{A} & \multicolumn{2}{c|}{B} &  \multicolumn{2}{c|}{C} \\
  \hline
  & \multicolumn{2}{c|}{$\chi^{2}_{\rm red}$ = 4.09} & \multicolumn{2}{c|}{$\chi^{2}_{\rm red}$ = 0.23} & \multicolumn{2}{c|}{$\chi^{2}_{\rm red}$ = 1.36} \\
  & \multicolumn{2}{c|}{$n_{\rm f} = 6$} & \multicolumn{2}{c|}{$n_{\rm f} = 3$} & \multicolumn{2}{c|}{$n_{\rm f} = 3$}\\
  & \multicolumn{2}{c|}{$\sigma_{\rm tot} = 0.37$} & \multicolumn{2}{c|}{$\sigma_{\rm tot} = 279.73$} & \multicolumn{2}{c|}{$\sigma_{\rm tot} = 140.28$} \\
  & \multicolumn{2}{c|}{$n_\mathrm{s}$ = 5662} & \multicolumn{2}{c|}{$n_\mathrm{s}$ = 3763} & \multicolumn{2}{c|}{$n_\mathrm{s}$ = 8031} \\
\hline
& Mean & Std & Mean & Std & Mean & Std \\
				\hline
$g_{1,1}$ & -0.29 & 0.04 & -0.18 & 0.06 & -0.13 & 0.07 \\
$g_{1,2}$ & -1.21 & 0.19 & 0.30 & 0.11 &0.27 & 0.10 \\
				\hline
$\mathcal{J}_{2}$ & -0.62 & 0.04 & -0.47 & 0.12 & -0.49 & 0.13 \\
$f_{2}$ & 0.27 & 0.03 & 1.40 & 1.30 & 1.15 & 0.74 \\
$g_{2,1}$ & -0.31 & 0.04 & 0.59 & 0.46 &0.65 & 0.30 \\
$g_{2,2}$ & -0.92 & 0.03 & -2.02 & 1.20 & -1.70 & 0.61 \\
				\hline
$\mathcal{J}_{3}$ & 0.38 & 0.04 & 0.77 & 0.14 & 0.71 & 0.13 \\
$f_{3}$ & 1.16 & 0.21 & 0.88 & 0.10 & 0.82 & 0.10 \\
$g_{3,1}$ & -0.66 & 0.35 & 0.18 & 0.10 &0.20 & 0.10 \\
$g_{3,2}$ & -1.67 & 0.68 & 0.26 & 0.14 & 0.28 & 0.14 \\
				\hline
$\mathcal{J}_{4}$ & 0.01 & 0.01 & -0.25 & 0.07 & -0.21 & 0.07 \\
$f_{4}$ & -0.03 & 0.02 & 1.39 & 171.91 & -0.64 & 108.71 \\
$g_{4,1}$ & -0.18 & 0.10 & 1.22 & 114.49 & -0.17 & 97.98 \\
$g_{4,2}$ & -1.01 & 0.05 & -2.69 & 261.05 & -0.16 & 150.61 \\
				\hline
$\mathcal{J}_{5}$ & -0.09 & 0.02 & -0.20 & 0.06 & -0.24 & 0.07 \\
$f_{5}$ & 0.13 & 0.03 & 2.29 & 241.46 & 1.28 & 119.32 \\
$g_{5,1}$ & -0.18 & 0.09 & -4.58 & 438.96 & -2.51 & 184.66 \\
$g_{5,2}$ & -0.93 & 0.02 & -0.88 & 171.57 &-0.90 & 77.17 \\
				\hline                                                          
			\end{tabular}
		\end{center}
\end{table}

\subsection{Comparison to previous lens-model-based results}
\label{sec:comparison_previous}

Next, we address the offset between light and mass for G1, as mostly inferred in previous works (see Table~\ref{tab:related_work}), but also of G2--4. 
\cite{bib:Massey2018} reached the conclusion that the offset between the centre of light of G1 and the peak of the mass density corresponding to G1 is below 2-$\sigma$ significance for their \lenstool and \grale reconstructions of the central mass density. 
They state an offset of 0.54~kpc corresponding to 0.29'' at the cluster redshift in a cosmology as observed by \cite{bib:Planck} (see also Table~\ref{tab:related_work}).
\news{This is a strong decrease in significance compared to the offset of about 2.1~kpc stated in \cite{bib:Mohammed} using the same \grale approach. 
Yet, \cite{bib:Mohammed} used less features per multiple image, in particular only two features in image~5 without having any information on image~4, nor the central images~6 and 7. 
Comparing the reconstructions in \cite{bib:Massey2015} and \cite{bib:Massey2018}, it is clear that the latter two images drive the lens models to represent the four-galaxy configuration in the cluster centre in the first place, as reconstructions without them hardly arrive at four mass peaks and yield large, significant offsets. 
This can be understood based on the findings summarised in \cite{bib:Wagner_summary} that these two images are necessary to constrain the mass-density profile in the central region. 
They thus mainly contribute to reduce offsets between mass and light for galaxies G3 and G4, being close to these two galaxies.
As further explained in \cite{bib:Wagner_frb}, a measured time-delay difference between a central image and any outer one would be able to alleviate the mass-sheet degeneracy and further increase the accuracy of the reconstruction.}

\news{Images~6 and 7 are, however, too far away from G1 to have an influence on its offset, which is also why adding additional galaxies in \cite{bib:Taylor} did not improve the reconstruction precision and accuracy in the vicinity of G1. 
The decrease in significance for this offset comes from the increasing amount of constraints by the additional features in images~4 and 5, as already found for the multiple-image configuration in CL0024 \citep{bib:Wagner_cluster}.}

\news{While the lens-model-driven offsets between light and mass in the central galaxies can thus be resolved by the lack of constraining data in their vicinity, the mass-density reconstructions deserve a second look to understand how they can generate the unusual cusp configuration of images~1--3 and the same-parity configuration of images~4 and 5. 
Taking into account that a mass-density distribution of four central masses can generate higher-order singularities as also elaborated in \cite{bib:Orban}, the lens models of \cite{bib:Massey2018} and \cite{bib:Chen} seem to fulfil all QFC to sufficiently explain the multiple-image configuration as an ``exotic'' single-plane lens. 
Concerning the local landscape of caustics and critical curves, the same-parity configuration of images~4 and 5 requires a kind of beak-to-beak caustic structure to produce the critical curve of two touching or almost touching closed loops as observed in the \glafic model by \cite{bib:Chen}. 
At the same time, using \grale instead of \glafic\!, this configuration causes an external mass agglomeration to be placed at the boundary of the reconstruction field in order to create the necessary external shear, as shown in \cite{bib:Massey2018}. 
This hypothesis is supported by comparing the two \grale reconstructions in \cite{bib:Mohammed} (see Fig.~5 of that work) with and without including arc~B as a constraint. 
The ``ghost mass'' south of image~4 is redistributed as soon as arc~B close to that clump is included as another image constraint. 
A similar construction to generate external shear can be observed in the \grale models of \cite{bib:Mohammed} above the cusp configuration.
Both of these external-shear masses still need to be physically motivated by complementary data. 
For instance, additional weak-lensing observables in the wider environment of A3827 could support or refute this interpretation.
An alternative interpretation to avoid additional external shear is outlined in Section~\ref{sec:multiple-lens-plane_extension}.
It shows that the unusual dynamics of four central galaxies and the close proximity of the cluster to us can also play a decisive role. 
}

\vspace{-2.5ex}
\subsection{Summary of contributions}
\label{sec:summary_contributions}

Based on all results obtained from analysing the anomalous multiple-image configuration of the source at redshift $z_\mathrm{s}=1.24$ behind A3827, we summarise our contributions to gain a deeper understanding of the mass density distribution in this extraordinary galaxy-cluster centre consisting of four galaxies G1--4 enclosed in the Einstein ring of the multiple-image configuration and a fifth galaxy G5 close by. 

We only use the single HST F336W filter band observation to identify the brightness features to be matched across all multiple images because this filter band shows the most features without being biased by foreground light from G1--4. 
In contrast to all previous works, we do not attempt to subtract the foreground light by using any surface brightness profile models for the galaxies. 
Instead, we directly apply our robust feature extraction approach as established in \cite{bib:Lin} to the observation and thereby avoid potential biasing in the identification of features based on the models for the intensity profiles of the galaxies. 

Since there are many objects at least 3-$\sigma$ above background noise level that could be matching features across all images, we are the first to systematically analyse the impact of the feature matching on the quality of the mass-density reconstruction. 
To do so, we set up the procedure summarised in Section~\ref{sec:joint_best-fit} to arrive at a self-consistent set of best-fit features and, based on the QFC that we establish, local lens properties at the positions of all multiple images as can be inferred from the lens-model-independent approach detailed in \cite{bib:Wagner_summary}. 
The method has already been applied to the multiple-image configuration in CL0024 \citep{bib:Wagner_cluster} and of Hamilton's Object \citep{bib:Griffiths, bib:Lin}.
Yet, both configurations showed a feature matching in agreement with standard lensing theory and also in accordance with the expectation of constant light-deflecting properties across the area covered by each multiple image, which is not the case in A3827. 
Consequently, even without imposing any assumptions about the \emph{global} light-deflecting mass density in terms of a lens model, we find in Section~\ref{sec:optimum_feature} that \emph{none} of the possible feature matchings yields physically plausible local lens properties. 
This means that it is unclear how the directions of the reduced shear and the ratios of the mass densities between the different image positions can be brought into accordance with expectations from standard lensing theory, other multiple-image configurations observed in other clusters, or from the observable foreground light, if light traces mass. 
From a comparison with lens-model-based approaches in previous works (see Table~\ref{tab:related_work}), it becomes obvious that some reconstructions hint at feature matchings close to the $\kappa=1$ isocontour.
Their inferred local lens properties are numerically unstable, so that the increasing confidence bounds cover any possible inconsistencies.

\comm{As Section~\ref{sec:offset_analysis} discusses, the local lens properties change over the area of the multiple images, so that estimating the local lens properties at the positions of the lensing galaxies cannot be obtained by extrapolating from those at the image positions.}
Taking all these results into account, we conclude that the previously claimed offset between light and mass in this cluster has been an artefact of the lens models in combination with the features and their matching chosen.
However, without a unique matching prescription and the possibility that features are occluded by the foreground light, it is impossible to agree on a specific value for the offset between light and mass, even assuming we could break the mass-sheet degeneracy and had enough central images to obtain an accurate lens model in the central part. 

We will elaborate on a plausible improvement of the modelling in Section~\ref{sec:multiple-lens-plane_extension}, as it can be motivated from the dynamical non-equilibrium state of the visible central galaxies and the anomalous feature matching across all multiple images that has evolved as the best one from our procedure in Section~\ref{sec:joint_best-fit}. 
Due to the strong bias in terms of foreground light and the sparsity and low resolution of complementary data available, there is still room for further ambiguities that could reconcile the observations with standard single-plane lens theory.
If a weak-lensing analysis can corroborate the external shear predicted by the \grale reconstructions, the \cite{bib:Massey2018} and \cite{bib:Chen} feature matchings could be supported. 
To shed further light on the offsets of the central galaxies, particularly G2--4, additional multiple images or complementary data are needed to fix the freedom in their mass peak positions.  
The results of these analyses can then be compared to the alternative interpretation as a thick lens sketched in Section~\ref{sec:multiple-lens-plane_extension}, which also shows a high degree of self-consistency for all data available in A3827. 
Any other explanation will need to achieve the same degree of self-consistency and the parameter space of possible models is greatly shrinking with an increasing amount of studies (see Table~\ref{tab:related_work}).

\section{Conclusion}
\label{sec:conclusion} 

In this work, we re-analysed the highly detailed multiple-image configuration in the galaxy-cluster strong gravitational lens A3827. 
Our main motivation was to resolve the contradicting conclusions that have been drawn on the potential offset between light and mass. 
As summarised and discussed in relation to all existing studies in Section~\ref{sec:comparative_discussion}, we corroborated the findings of \cite{bib:Massey2018} that the previously claimed offset at about 3-$\sigma$ significance is not physical but caused by insufficient modelling of the light-deflecting mass-density distribution. 

The cluster is neither in dynamical equilibrium, nor does the large extent of foreground light allow for a unique identification of the required details within all extended multiple images.
After having excluded microlensing, source-intrinsic transient phenomena, and potential contaminations with small foreground objects, we found that the multiple-image configuration, as characterised by the identification of brightness features across all images, deviates from standard ones in single-lens-plane scenarios.
We therefore suggest an extension to at least two lens planes as detailed in Section~\ref{sec:multiple-lens-plane_extension}. 
As discussed, this idea is physically plausible because it shows consistency to the strong-lensing and complementary data.
It also avoids the additional external shear required to explain the unusual cusp configuration for images~1--3 and could facilitate the explanation of the same-parity configuration for image~4 and 5, whose origin as a higher-order singularity still needs to be derived from the strong lensing formalism. 
 
In addition, we briefly comment on the usefulness of this data set to probe dark-matter models beyond cold dark matter or test alternative gravity models like proposed in \cite{bib:Chen} in Section~\ref{sec:sidm} before we conclude with a general outlook how to tackle upcoming anomalous configurations like A3827 in Section~\ref{sec:outlook}.

\subsection{Multiple-lens-plane extension}
\label{sec:multiple-lens-plane_extension}

\begin{figure}
	\centering
	\includegraphics[width=0.47\textwidth]{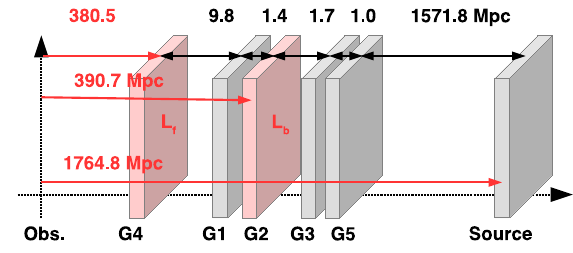}
	\caption{Multiple-lens-plane scenario for A3827 based on the redshifts and their uncertainties of \protect\cite{bib:Massey2015}. All angular diameter distances are based on \protect\cite{bib:Planck}. Uncertainties in redshifts allow G1 and G2 to lie in a single lens plane as well as G3 and G5. Based on the bi-modal velocity distribution of member galaxies \protect\citep{bib:Carrasco}, G4 is assumed to lie in the foreground plane L$_\mathrm{f}$, all other galaxies are assumed to be in a background plane L$_\mathrm{b}$ in the plane of G2.}
	\label{fig:multi-plane_lens}
\end{figure}

As shown in all previous sections, how this anomalous multiple-image configuration came about is still not answered sufficiently because it is in contradiction with our expectations from single-lens plane standard lensing theory \citep{bib:SEF, bib:Petters}. 
We therefore follow up on the result noted in \cite{bib:Carrasco} that the kinematical analysis of the member galaxies of A3827 shows a bi-modal distribution of velocities.
Hinting at a merger event, this observation suggests a re-analysis of the multiple-image configuration using at least two lens planes. 
Calculating the angular diameter distances of galaxies G1--5 based on the cosmology by \cite{bib:Planck} for the redshifts measured in \cite{bib:Massey2015} (see their Table~1), we indeed find support for this hypothesis, as shown in Fig.~\ref{fig:multi-plane_lens}. 
The uncertainties in the measured redshifts allow for G1 and G2 to be at the same distance, G3 is slightly behind that and can be at the same distance as G5. 
G4 is about 10~Mpc in front of these galaxies opening a separate lens plane. 
We thus set up a first crude division of lens planes into L$_\mathrm{f}$ containing galaxy G4, being located at its distance $D_\mathrm{f}=D(z_4)$ and L$_\mathrm{b}$ containing the other galaxies and, for the sake of simplicity, being located at the distance of G2, $D_\mathrm{b} = D(z_2)$.  
Assuming that both lens planes contained all light-deflecting mass compressed to a point-mass, $M_\mathrm{f}$ and $M_\mathrm{b}$, respectively, we can constrain their ratio of (angular) Einstein radii $\theta_\mathrm{E}$ as 
\begin{equation}
\eta_\mathrm{E} \equiv \dfrac{\theta_\mathrm{E,f}}{\theta_\mathrm{E,b}} = \sqrt{\dfrac{M_f}{M_b}} \;\sqrt{\dfrac{D_\mathrm{fs}}{D_\mathrm{f}}\dfrac{D_\mathrm{b}}{D_\mathrm{bs}}}  \;.
\label{eq:thetaE}
\end{equation}

Using the masses inferred for the stellar part of the galaxies G1--5 and the total intra-cluster gas from \cite{bib:Chen} (see their Table~2) and assuming that these luminous masses are all scaled by roughly the same factor to account for their dark-matter part, the latter cancels out in the mass ratio of Eq.~\ref{eq:thetaE}. 
We can thus calculate the estimate for the total mass ratio between L$_\mathrm{f}$ and L$_\mathrm{b}$ by assuming that half of the intra-cluster light and G4 constitute the luminous mass of L$_\mathrm{f}$, $M_\mathrm{f}=0.5 M_\mathrm{gas}+M^{*}_\mathrm{G4}$, and that half of the intra-cluster light and the stellar mass of G1, G2, G3, and G5 form the luminous mass of L$_\mathrm{b}$. 
Inserting the distance and luminous mass ratios into Eq.~\ref{eq:thetaE}, we obtain $\eta_\mathrm{E} = 0.81$.

\begin{figure}
	\centering
	\includegraphics[width=0.4\textwidth]{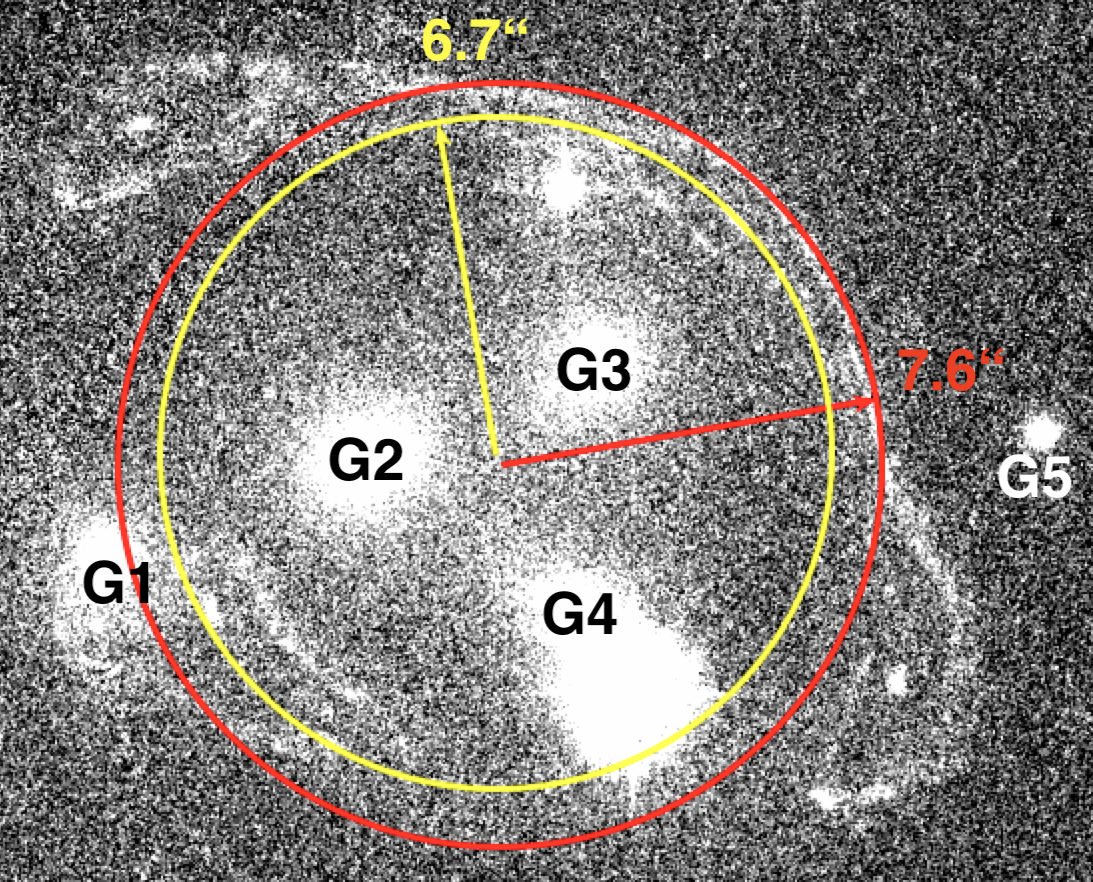}
	\caption{Visual fit of two Einstein radii (yellow and red circles) to the observable multiple-image configuration, corresponding to the approximation of two lens planes L$_\mathrm{f}$ and L$_\mathrm{b}$ containing point masses $M_\mathrm{f}$ and $M_\mathrm{b}$, respectively (see Fig.~\ref{fig:multi-plane_lens}). Central cluster member galaxies are marked as G1--5.}
	\label{fig:circle_fit}
\end{figure}

To estimate the same ratio from the observables in the F336W filter band, we assume that there are two ring-like brightness structures visible close to image~2 to which we visually fit circles, as sketched in Fig.~\ref{fig:circle_fit}. 
Reading off the radii of these circles that could be the two Einstein rings for the two lens planes, $\theta_\mathrm{E,1}=7.6''$ (red circle) and $\theta_\mathrm{E,2}=6.7"$ (yellow circle), we arrive at $\eta_\mathrm{E}=0.88$, which is close to the crude estimate obtained from the complementary data. 
Taking a closer look at the multiple images, the spiral arms appear defocussed and look ``washed out'' in the observation, supporting the hypothesis of two lens planes and a cumulative lensing effect \citep{bib:Petters}. 

\news{Most importantly, the anomalous relative orientation between images~1--3 can either be explained as a sequence of shearing and scaling (as clearly observed in the \grale reconstructions in \cite{bib:Mohammed}).
For the anomalous configuration of images~4 and 5, \cite{bib:Massey2018} reconstructions show external mass agglomerations creating the necessary shear in a similar way.}
Yet, it is also possible to reduce this sequence to an effective shear and a rotation. 
The latter cannot be caused by an individual lens plane.
Therefore, it seems more plausible to interpret the anomalous configuration in the simpler terms of an effective shear and a rotation caused by a multi-plane scenario instead of being forced to add many mass clumps at positions where no luminous counterpart may be found to generate the configuration observed with single-lens-plane effects only \news{(compare the larger environment of A3827 beyond the region used in the lens-model reconstructions with the external shear directions of the \grale reconstructions for a first estimate)}. 

Transferring these insights into galaxy-scale lensing, for which line-of-sight effects are more abundant than for galaxy-cluster-scale lenses, it is possible that the discrepancies between modelled external shear and external shear as inferred from weak-lensing observations found in \cite{bib:Etherington} are of the same multi-plane origin. 
Thus, follow-up analyses taking account of multiple lens planes are in order for these galaxies. 
Contrary to that, galaxy clusters at higher redshift may still be modelled by the thin-lens approximation, as multi-plane effects for A3827 may only arise due to its very low redshift and the fact that the foreground-lens-plane distance to the background one amounts to 3\% of the distance between the background lens plane and us.

\subsection{Self-interacting dark matter extension and alternative gravity tests}
\label{sec:sidm}

As started by the large offset between light and mass for G1 inferred in \cite{bib:Williams}, A3827 has been considered as a test bed for alternative types of dark matter beyond the standard cold, collisionless dark matter and most recently as a test for alternative theories of gravity like the MOdified Newtonian Dynamics (MOND) (see \cite{bib:Milgrom} for a recent review).
However, the dynamically non-relaxed structure of A3827 and the complication of having to solve an at least four-body problem in the cluster centre make it difficult for both theoretical frameworks to distinguish an oversimplified modelling not accounting for the kinematics in a proper way from actual hints of modifications of gravity or other forms of dark matter, like self-interacting dark matter (SIDM).

While these issues were raised in previous works, complementary studies explored the possibility of constraining SIDM models \citep{bib:Kahlhoefer, bib:Schaller} to improve on the simple one for a lower bound on the SIDM cross section in \cite{bib:Williams}. 
\cite{bib:Schaller} characterised the significance of the offset between light and mass in G1 as constrained by \cite{bib:Massey2015} in terms of a comparison to the \eagle simulation and concluded that it also was at odds with the simulated universe at 3-$\sigma$. 
Yet, follow-up simulations at higher resolution and in larger volumes are necessary to gain an understanding of the origin and the evolution of this offset and to strengthen the statistical significance of this comparison.
In particular, a statement on the oddity of such an offset should be made after having investigated the probability of occurrence of such a configuration of central galaxies in a cosmic structure.

Similarly, \cite{bib:Kahlhoefer} stated that the inferred offset in \cite{bib:Massey2015} already excluded many types of dark matter candidates like axions, neutrinos, and ``weakly-interacting massive particles''.
The constraint of self-interaction from A3827 was therefore considered quite tight for astro-particle physics and in tension with constraints found in other astrophysical systems, as argued by the authors. 
Their modelling also inspired the establishment of the tilted mass density profile employed in \cite{bib:Taylor} as a second option to study the displacement between light and mass in terms of skewness because \cite{bib:Kahlhoefer} found that the offset depended on the definition of the mass clumps, their centroid and peak positions (see a further discussion on the emergence of structures and the problem to define their boundaries in \cite{bib:Wagner_daemon}). 

\cite{bib:Chen} suggested using A3827 to compare a dark-matter-based lens reconstruction with alternative theories of gravity.
In MOND, for instance, luminous masses along the line of sight contribute to the lensing effect, so that light deflection does not only happen in a single lens plane \citep{bib:Mortlock, bib:Milgrom}. 
Consequently, as they argue, estimating the acceleration at the radius of the multiple-image configuration to be above the MONDian acceleration scale hints at the lens to be thin and be above the scale of relevance for MOND. 
This, however, neglects the external field effects \citep{bib:Milgrom} typical of MOND which can occur due to the non-linearity of the modified Newtonian acceleration, such that gravitational interactions of subsystems do not decouple from their environment. 
For the unknown dynamical evolution and state of A3827, it may play a role and mimic an acceleration within the Newtonian regime. 
Based on the prerequisite of a thin lens, \cite{bib:Chen} then employ the concept of phantom dark matter (see, for instance, \cite{bib:Milgrom_pdm} for details and \cite{bib:Oria} an application) to investigate the equivalent total Newtonian mass that should trace the visible mass if it is supposed to represent the total (only luminous) mass density in a MONDian theory of gravity. 
While the concept of phantom dark matter is definitely helpful to compare MOND with dark-matter scenarios, it also needs to take into account the multi-plane structure of the central galaxy distribution detailed in Section~\ref{sec:multiple-lens-plane_extension}.

In summary, follow-up studies to find explanations for the offset in terms of new dark-matter properties quickly produced inconsistencies with existing candidate models or simulations. 
In future anomalous observations, such hints should hence be taken as a sign to cross-check the model assumptions generating the anomaly.
For instance, \cite{bib:Massey2015} already noted that the offset depends on the location of the brightness features used in the lens modelling. 
For A3827, we can clearly point at the ambiguous identification of brightness features and at the difference between data-inferred \emph{local} lens properties that can be directly constrained by observables and are only valid at the positions of the multiple images and the estimates of lens properties farther away from the data points which are model-driven \citep{bib:Wagner_summary}. 
Furthermore, the test of alternative gravity theories as proposed in \cite{bib:Chen} cannot be applied to A3827 due to its dynamical non-equilibrium state, in particular its potential deviation from a thin lens (as outlined in Section~\ref{sec:multiple-lens-plane_extension}). 
Testing MOND as one example, the theory to describe galaxy-cluster lenses is still in its development and there is a degeneracy between external field effects and a standard Newtonian acceleration in such dynamical environments increasing the difficulty to draw significant conclusions.
As the offset was inferred from insufficient modelling and constraints on its existence turned out to be inconclusive, it is thus not possible to refute MOND on the basis of this argument.

\subsection{Outlook}
\label{sec:outlook}

On the whole, we showed that highly detailed multiple-image configurations lack the precision and accuracy to constrain the total mass density distribution for individual lensing galaxies, even if the multiple images are as close to the galaxy as images~4 and 5 are to G1 in A3827 (see also \cite{bib:Wagner_cluster, bib:Griffiths}, and \cite{bib:Lin} for further details).
On the other hand, we could demonstrate that the details observable in the very same configurations already have the precision and accuracy to require lens reconstructions beyond the effective single-lens-plane approach.
For A3827, this could have been expected from the complementary spectroscopic data obtained in the first work \citep{bib:Carrasco}, revealing an at least bi-modal merging structure along the line of sight and given that the extension of this merger occupied a significant portion of the total distance from the observer to the lens. 
Whether or not a multi-plane lens represents the data in a more realistic way can also be investigated by adding weak-lensing measurements to the single-plane reconstructions. 
The latter scenario yields a plausible explanation, if the required external shear in the single-plane scenario can be attributed to physically existing or plausible \comm{external masses \citep{bib:Etherington, bib:Fleury3, bib:Hogg}}. 
\news{Based on strong-lensing observables alone, the distribution of dark matter in lens-model-based mass-density reconstructions can thus suffer from a full three-dimensional degeneracy and the implications of this degeneracy, for instance on approaches as outlined in \cite{bib:Cerini}, require further analyses.}

A major lesson to be learned from A3827 is that all results obtained for the cluster clearly hinted at an oversimplified modelling.
There were many complementary inconsistencies emerging in follow-up studies about the consequences of the interpretation, as further detailed in Sections~\ref{sec:sidm} and \ref{sec:multiple-lens-plane_extension}, respectively. 
Thus, exploring unprecedented, exotic new physical implications should not be considered until all simpler options using known astrophysics and exploring the limits of all approximations and assumptions used have been cross-checked for their continuing validity.

\vspace{-1.5ex}
\section*{Acknowledgements}

We would like to thank Pavel Kroupa, Jori Liesenborgs, Richard Massey, Subir Sarkar, \comm{and our referee} for helpful comments and discussion, Nicolas Tessore for his image mapping software, \ptmatch\!, Kyle Barbary for his Python code of SExtractor, \sep\!, which both were crucial to this project.
This paper uses data from the Hubble Legacy Archive, which is a collaboration between the Space Telescope Science Institute (STScI/NASA), the Space Telescope European Coordinating Facility (ST-ECF/ESAC/ESA) and the Canadian Astronomy Data Centre (CADC/NRC/CSA).

\vspace{-2ex}
\section{Data availability}

The observation of the HST F336W filter band used in this work is publicly available in the Hubble Legacy Archive, all sources of data in original and processed form are listed in Table~\ref{tab:filter_bands}. 
\sep\!, and \ptmatch\! are open source software which can be downloaded here:
\url{https://sep.readthedocs.io/en/v1.1.x/}, and \url{https://github.com/ntessore/imagemap}.
Our own code for the persistence pipeline and the evaluation of the \ptmatch results is available here:
\url{https://github.com/joycelin1123/SEP-Automated-Feature-Detection}. 
 

\bibliographystyle{mnras}
\bibliography{references} 


\appendix

\section{Supplements for brightness feature extraction}
\label{app:brightness_feature}

\begin{figure*}
\centering
\includegraphics[width=0.45\textwidth]{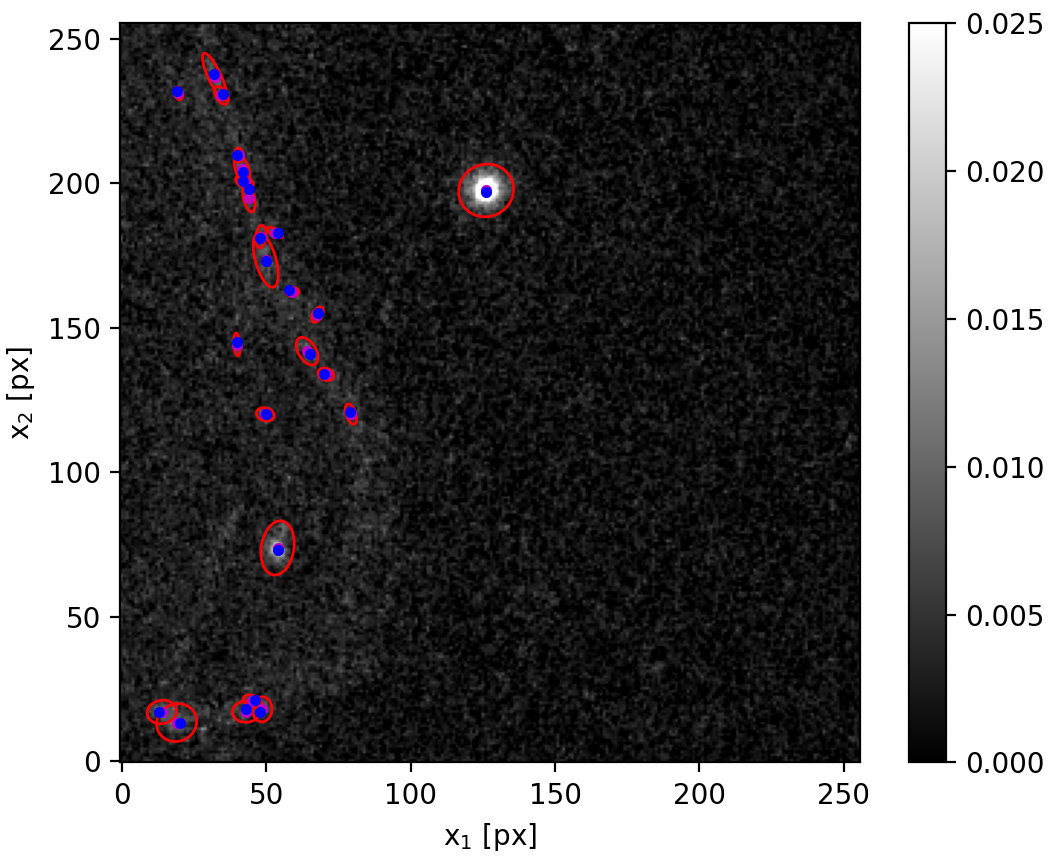} \hfill
\includegraphics[width=0.45\textwidth]{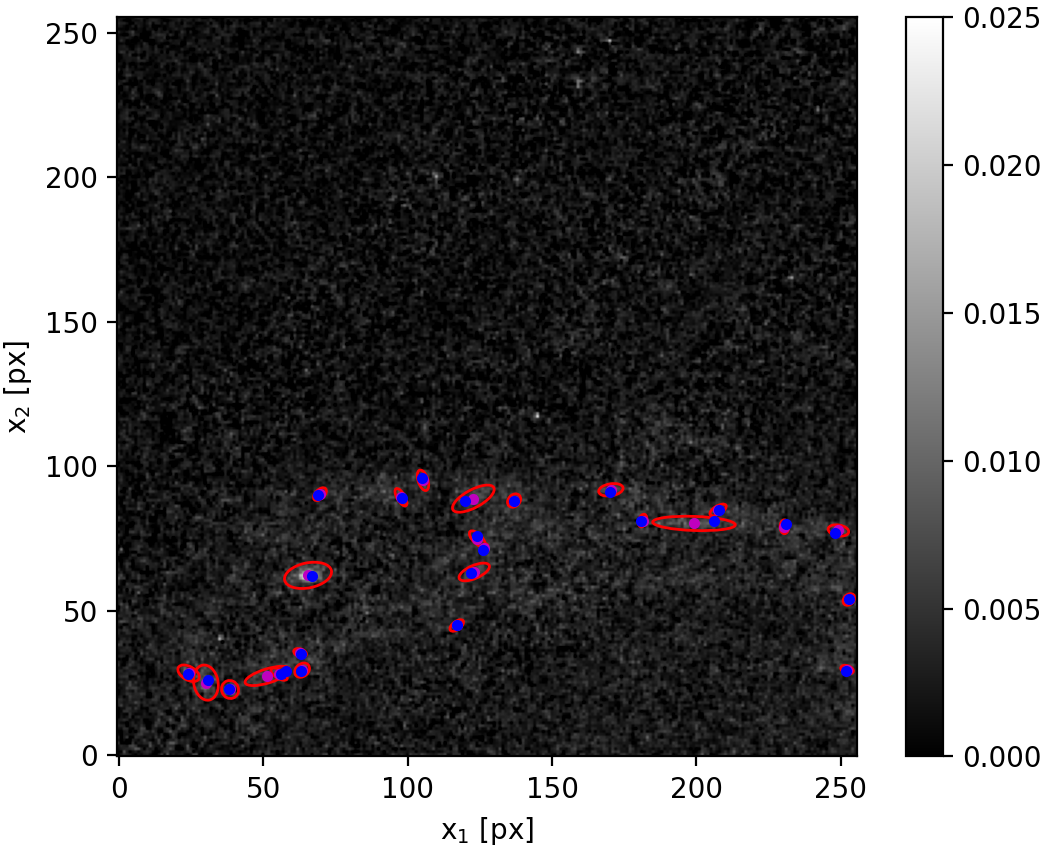}
\caption{Same as Fig.~\ref{fig:image245_objects} for image~1 (left) and 3 (right).}
\label{fig:image13_objects}
\end{figure*}

\begin{figure*}
\centering
\includegraphics[width=0.43\textwidth]{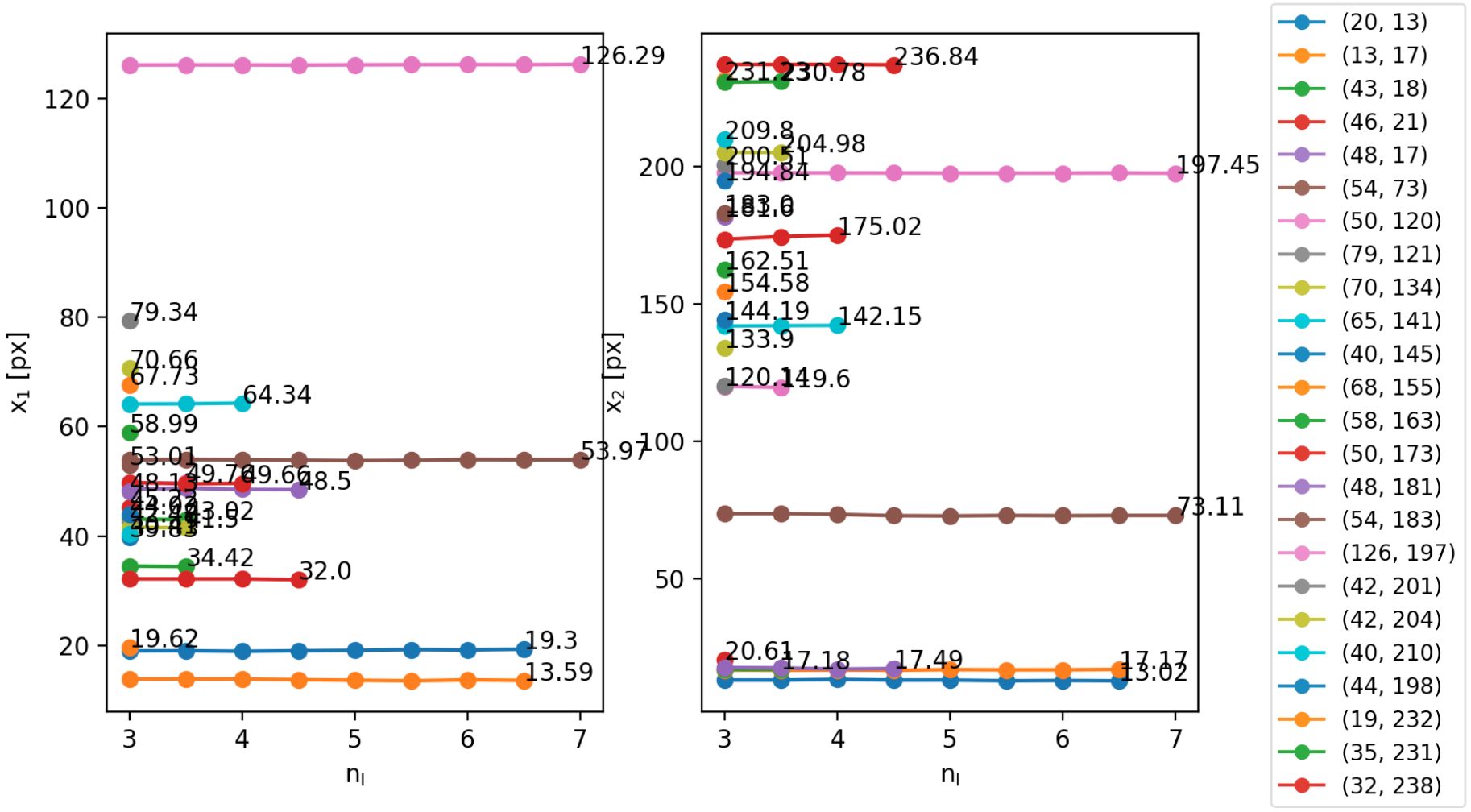} \hfill
\includegraphics[width=0.54\textwidth]{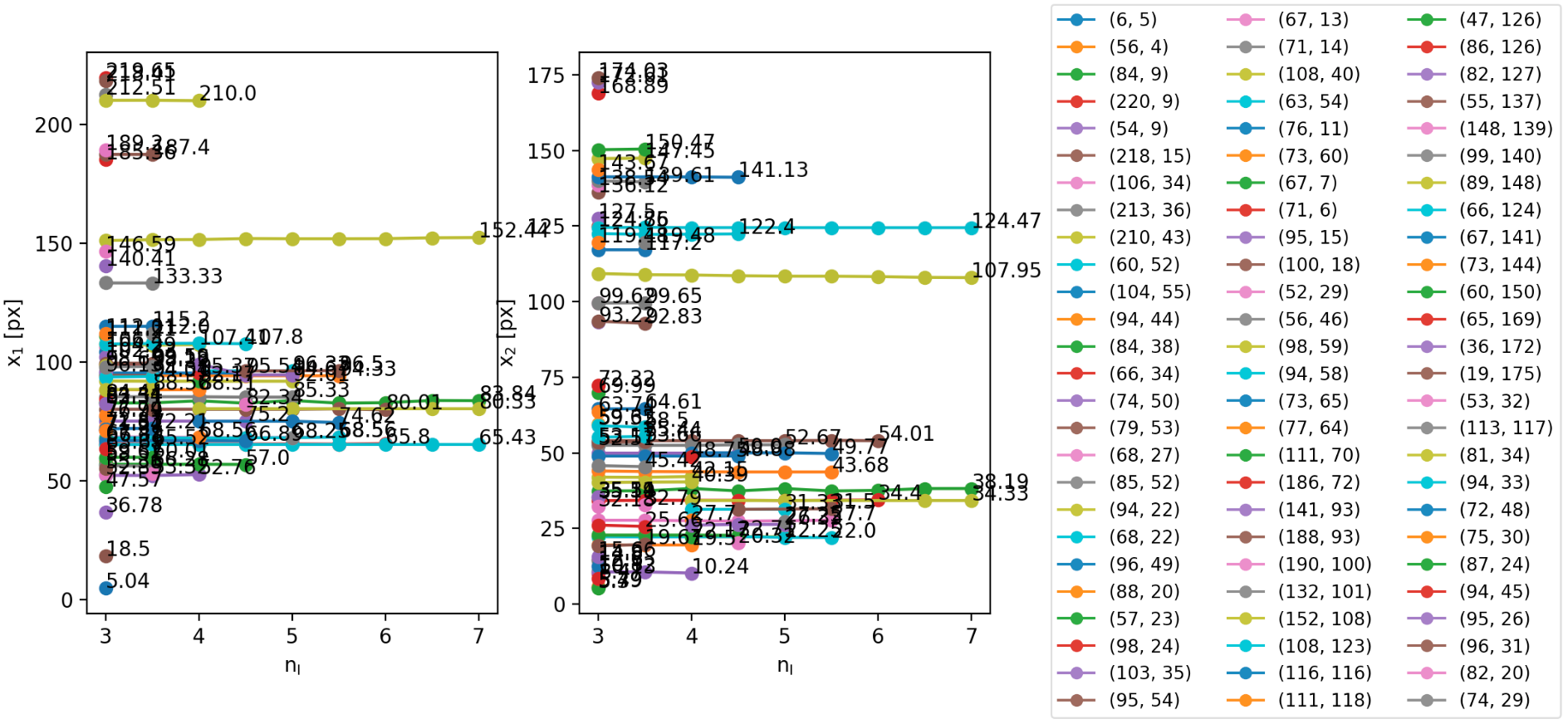}
\caption{Persistence diagram for objects detected in image~1 (left) and 2 (right). The centre of light positions are plotted as $x_1$ (left-hand plot for each image) and $x_2$ (right-hand plot for each image) over increasing threshold in intensity $n_\mathrm{I}$. Coordinates $(x_1,x_2)$ according to Figs.~\ref{fig:image13_objects} (left) and \ref{fig:image245_objects} (left).}
\label{fig:image12_persistence}
\end{figure*}

\begin{figure*}
\centering
\includegraphics[width=0.44\textwidth]{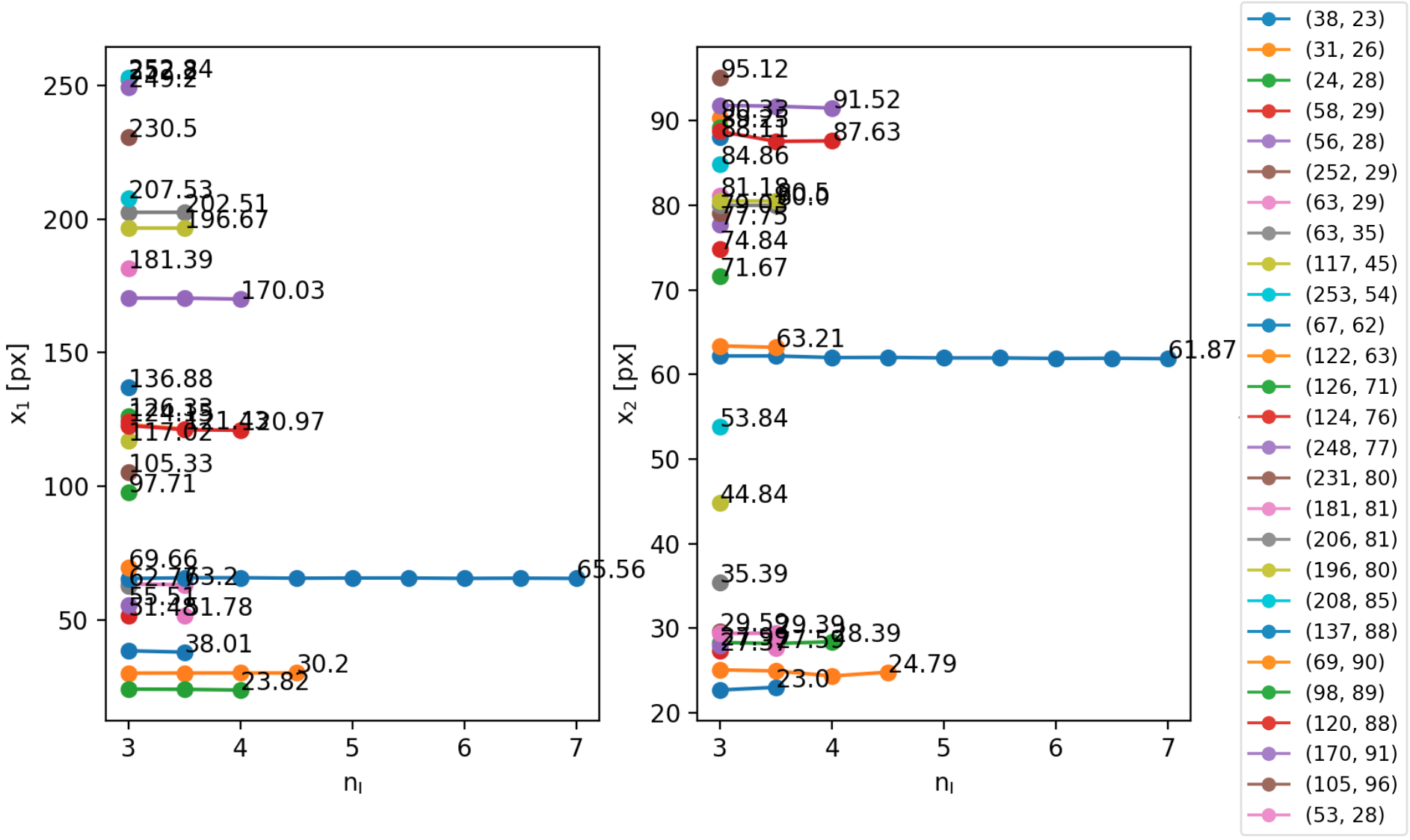} \hfill
\includegraphics[width=0.54\textwidth]{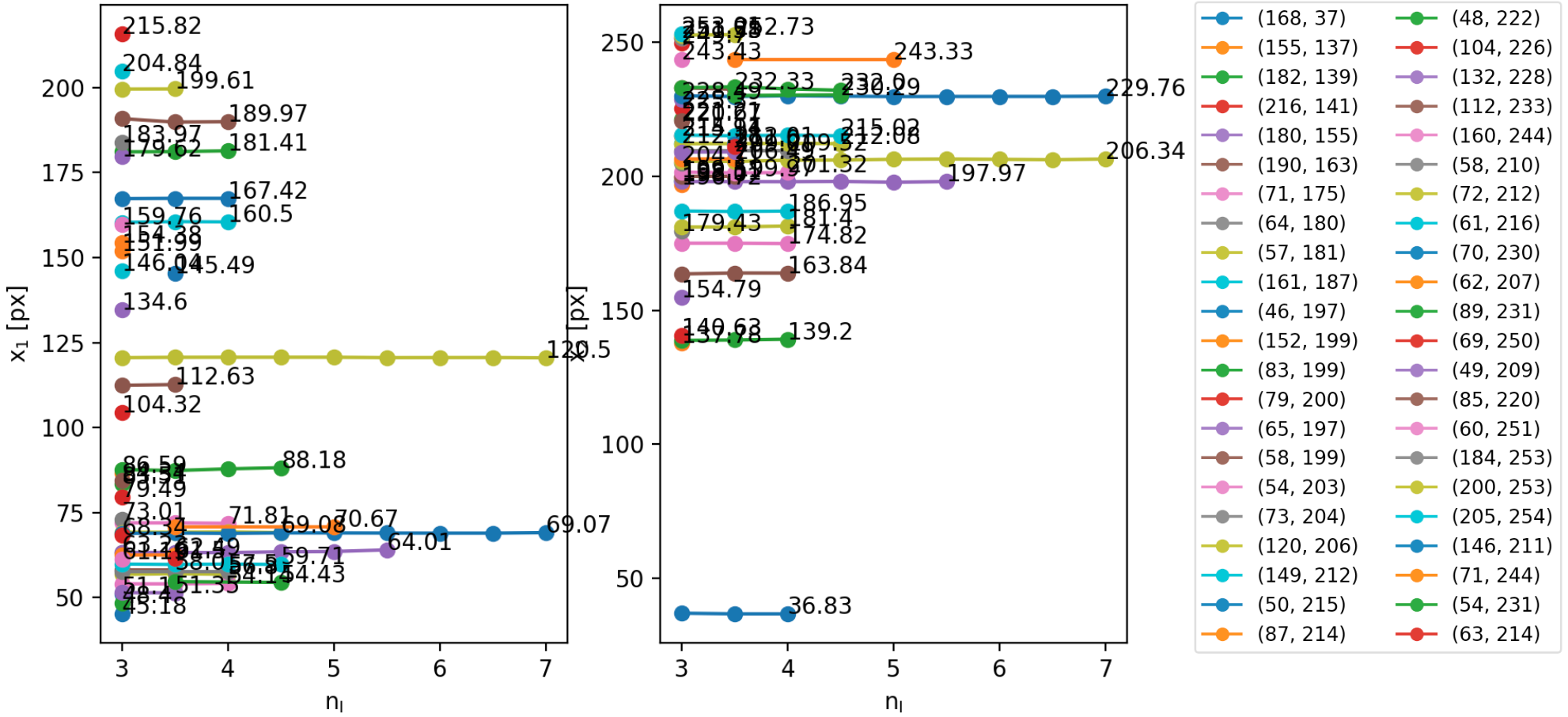}
\caption{Same as Fig.~\ref{fig:image12_persistence} for image~3 (left) and images~4 and 5 (right).}
\label{fig:image345_persistence}
\end{figure*}

As stated in Section~\ref{sec:brightness_feature}, we show the remaining plots for the object detection in images~1 and 3 in Fig.~\ref{fig:image13_objects}, as well as the persistence diagrams of images~1--5 in Figs.~\ref{fig:image12_persistence} and \ref{fig:image345_persistence}.

\bsp	
\label{lastpage}
\end{document}